\documentclass[preprint]{aastex}

\usepackage{graphicx}
\usepackage{natbib}

\newcommand{\msun}{{\,\rm M}_{\odot}}
\newcommand{\lsun}{{\,\rm L}_{\odot}}

\slugcomment{Draft version  from \today}

\shorttitle{CCH in DM~Tau, LkCa~15, and MWC~480}
\shortauthors{Henning et al.}

\begin{document}

\title{Chemistry in Disks. \\
III. -- Photochemistry and X-ray driven chemistry probed by the ethynyl radical
(CCH) in DM~Tau, LkCa~15,
and
MWC~480.}

\author{Th.~Henning and D. Semenov} \affil{Max-Planck-Institut f\"ur
  Astronomie, K\"onigstuhl 17, 69117 Heidelberg, Germany}
\email{henning,semenov@mpia.de}

\author{St. Guilloteau, A. Dutrey, F. Hersant, and V. Wakelam}
\affil{Universit\'e Bordeaux 1; Laboratoire d'Astrophysique de
  Bordeaux (LAB) and CNRS/INSU - UMR5804 ; BP 89, France}

\author{E. Chapillon} \affil{Max-Planck-Institut f\"ur
  Radioastronomie, Auf dem H\"ugel 69, 53121 Bonn, Germany}

\author{R. Launhardt} \affil{Max-Planck-Institut f\"ur Astronomie,
  K\"onigstuhl 17, 69117 Heidelberg, Germany}

\author{V. Pi\'etu}
\affil{IRAM, 300 rue de la piscine, F-38406 Saint Martin d'H\`eres, France}

\and

\author{K.Schreyer}
\affil{Astrophysikalisches Institut und Universit\"ats-Sternwarte,
Schillerg\"asschen 2-3, 07745 Jena, Germany}

\begin{abstract}
We studied several representative circumstellar disks surrounding the
Herbig Ae star MWC~480 and the T Tauri stars LkCa~15 and DM~Tau at
(sub-)millimeter wavelengths in lines of CCH\footnote{Based on
observations carried out with the IRAM Plateau de Bure
Interferometer. IRAM is supported by INSU/CNRS (France), MPG
(Germany) and IGN (Spain).}. Our aim is to characterize
photochemistry in the heavily UV-irradiated MWC~480
disk and compare the results to the disks around cooler T Tauri stars.
We detected and mapped CCH in these disks with the IRAM Plateau de Bure
Interferometer in the C- and D-configurations in the (1-0) and (2-1)
transitions. Using an iterative minimization technique, the CCH
column densities and excitation conditions are constrained. Very low excitation
temperatures are derived
for the T Tauri stars. These values are compared with the results
of advanced chemical modeling, which is based on a steady-state flared disk
structure with a vertical temperature gradient, and a gas-grain chemical network
with surface reactions. Both model and observations suggest that CCH is a
sensitive tracer of the
X-ray and UV irradiation. The predicted radial dependency and source to source
variations of CCH column densities
qualitatively agree with the observed values, but the predicted column densities
are too low by a factor of several. The chemical model fails to reproduce
high concentrations of CCH in very cold disk midplane as derived from
the observed low excitation condition for both the (1-0) and (2-1) transitions.
\end{abstract}

\keywords{stars: formation --- planetary systems: protoplanetary disks
  -- astrochemistry -- molecular processes -- stars: individual (DM
  Tau, LkCa 15, MWC 480)}

\section{Introduction}
Our present understanding of the chemical composition and evolution of
protoplanetary disks is far from being complete. Recent years have
revealed increased interest in such studies as a necessary
prerequisite to characterize the physical and chemical conditions for
planet formation. Apart from CO and its isotopologues, and
occasionally HCO$^+$, CN, HCN, and CS, the molecular content of
protoplanetary disks characterized by millimeter line observations
remains poorly known \citep[see reviews
  by][]{DGH07,Bergin_ea07}. Recently, infrared spectroscopy provided evidence
for the presence of C$_2$H$_2$, HCN, H$_2$O, OH, and CO$_2$ in
the inner regions of protoplanetary disks and indicated the importance
of photochemistry
\citep[e.g.,][]{Lahuis_ea06,Carr_Najita08,Salyk_ea08,Pascucci_ea09}.
Molecular line data are limited in their
sensitivity and spatial resolution, which implies that the spatial distribution
of molecular abundances in disks remains poorly determined
\citep[e.g.,][]{Pietu_ea05,Semenov_ea05,Qi_ea06,Qi_ea08,Panic_ea09}. Thus
a detailed comparison with existing chemical models is difficult and often
based on global data (e.g. integrated line profiles), while for testing the
validity of sophisticated models one has to include the thermal disk structure
and the time-dependent chemical evolution together with line data and channel
maps.

As part of the Heidelberg-Bordeaux ``Chemistry In Disks'' (CID)
project we are investigating the molecular content and physical properties of
a sample of well-studied T~Tauri and Herbig~Ae disks of various age,
followed by comprehensive physico-chemical modeling. In the first CID
paper, we have presented the results of a deep search for N$_2$H$^+$
and HCO$^+$ toward two T Tauri stars (DM~Tau, LkCa~15) and one Herbig
Ae star (MWC~480), see \citet[CID1 paper hereafter]{Dutrey_ea07}. The
N$_2$H$^+$ emission has been detected in LkCa 15 and DM Tau, with the
N$_2$H$^+$ to HCO$^+$ ratio of a few percent, similar to that of cold
dense cores, and the disk ionization degree of $\sim 10^{-8}$, as
predicted by chemical models.

In the second CID paper by \citet[]{Schreyer_ea08},
we have found that the modeled and observed column densities of
several key species relative to $^{13}$CO are lower for the Herbig AB
Aur disk than the values measured in DM Tau, while the absolute
amount of CO gas is similar in both disks. This has been
interpreted as an indication of a poor molecular content of the Herbig
A0e system compared to the disk around the M1e star DM Tau due to intense
UV irradiation of the disk by the A0 star.

In this third CID paper, we report on the observations of CCH in three
objects: DM~Tau, LkCa~15 and MWC~480. The CCH molecule has been
selected because it should be sensitive to the UV radiation field and
because its chemistry is relatively well-studied, with many accurately
determined relevant reaction rates \citep[e.g.,][]{Woodall_ea07}.
Using the $\chi^2$-minimization
technique in the $uv$-plane as described in \citet{GD98}, we derive
the column density of CCH in the upper layers of the outer disks. We
compare these values with the CCH column densities computed
with realistic two-dimensional steady-state disk models and a
gas-grain chemistry with surface reactions.

\section{Observations}
\label{obs}

\subsection{Sample of Stars}
The source properties and their coordinates are summarized in
Table~\ref{tab:coord}. All these systems are isolated and located in
regions devoid of CO contamination. The disk sizes, masses, and
ages are presented in Table~\ref{tab:disk_struc} (see also Tables~1
and 2 in CID1).

\begin{deluxetable}{llllllllll}
\tabletypesize{\scriptsize}
\tablecaption{Properties of the targets\label{tab:coord}}
\tablewidth{0pt}
\rotate
\tablehead{
\colhead{Source} & \colhead{Right Ascension} & \colhead{Declination} &
\colhead{Spect.Type} & \colhead{Effective Temp.} & \colhead{Stellar Lum.} &
\colhead{Stellar Mass} & \colhead{Age} & \colhead{UV flux} & \colhead{X-ray
lum.}\\
\colhead{} & \colhead{($^{o}$,$^{'}$,$^{''}$), J2000.0} &
\colhead{($^h$,$^m$,$^s$), J2000.0} & \colhead{} & \colhead{(K)} &
\colhead{($\lsun$)}  & \colhead{($\msun$)} & \colhead{(Myr)} &
\colhead{($\chi_0$)} & \colhead{erg\,s$^{-1}$}
}
\startdata
DM~Tau   & 04:33:48.73 & 18:10:09.89  & M1 & 3720 & 0.25 & $0.55\pm0.07$ & 5
&  300 & $3\,10^{29}$ \\
LkCa~15  & 04:39:17.78 & 22:21:03.34  & K5 & 4350 & 0.74 & $0.97\pm0.03$ & 3-5
& 1850 & $3\,10^{29}$ \\
 MWC~480  & 04:58:46.26 & 29:50:36.87  & A4 & 8460 & 11.5 & $1.65\pm0.07$ & 7
& 5000  & $10^{28}$\\
\enddata
\tablecomments{Coordinates J~2000.0 deduced from the fit of the 1.3mm
PdBI continuum map \citep{Pietu_ea06}. Col.3, 4, 5, 6 and 7 are the
spectral type, effective temperature, the stellar luminosity, the stellar mass,
and age as given in \citet{Simon_ea00}. The stellar FUV luminosities at
$r=100$~AU are given in Col. 8 in units of the \citet{G} interstellar UV field.
These values are rescaled from \citet{Bergin_ea04} (LkCa~15 and DM~Tau) for the
$1\,100-2\,070\AA$ region, assuming that $30\%$ of the FUV flux is
the Ly$_\alpha$ radiation. For MWC~480 the UV flux $\chi$ is computed from the
\citet{kurucz} ATLAS9 of stellar spectra. The stellar X-ray luminosities are
taken from \citet{Glassgold_ea05}.}
\end{deluxetable}

\begin{deluxetable}{lcccccc}
\tablecaption{Properties of protoplanetary disks\label{tab:disk_struc}}
\tablehead{
\colhead{Source}   & \colhead{$i$} & \colhead{R$_{out}$} & \colhead{$T_{100}$} &
\colhead{$q$} & \colhead{Acc. Rate} & \colhead{Disk Mass} \\
\colhead{} & \colhead{($^o$)} & \colhead{(AU)}  & \colhead{(K)} & &
\colhead{($\msun$\,yr$^{-1}$)} & \colhead{($\msun$)}
}
\startdata
DM~Tau   & -32 & 800 & 15 &   $0.12-0.5$ & 2.10$^{-9}$ & 0.05 \\
LkCa~15  &  52 & 680  & 15 &   $0.4$ & 10$^{-8}$   & 0.03 \\
MWC~480  &  38 & 695& 30 & $0.28-0.65$ & 10$^{-8}$   & 0.03 \\
\enddata
\tablecomments{Col.2 Inclinations are taken from
\citet{Simon_ea00}. Col.3 \& 4 Apparent disk radii and
temperature distribution are taken from the CO and HCO$^+$
interferometric analysis of \citet{Pietu_ea07}. Here $T_{100}$ denotes gas
kinetic temperature at 100~AU, $q$ - the power-law exponent in the temperature
distribution. Col.5 Accretion rates are taken as
a generic value from \citet{DAlessio_ea99} for MWC~480 and
LkCa~15. The accretion rate for DM~Tau is taken from
\citet{Calvet_ea05}. Col.6 Disk masses are from
\citet{Dutrey_ea97} for DM~Tau, \citet{Thi_ea04} for
LkCa~15 and \citet{Mannings_ea97} for MWC~480.}
\end{deluxetable}

\subsection{PdBI observations and data reduction}
\label{obs_pdbi}
The Plateau de Bure observations of the C$_2$H(1-0) line in DM~Tau were carried
out in September \& October 1997, and between October 2002 and March 2003,
mostly in compact configurations. The total observing time on source (equivalent
to 6 antennas) is about 30~hours and the angular resolution (with natural
weighting) is $3.5''$. The (1-0) line in LkCa~15 and MWC~480 was
observed in track-sharing mode in December 2004 and April 2005, with a total
observing time on source of  $\sim 5$~hours and an angular resolution of
$3.8''\times2.5''$.
The C$_2$H(2-1) line was observed in track-sharing mode for the three
sources in D configuration in August \& September 2008. The total
observing time on source with dual polarization
receivers is 5~hours. The angular resolution is $2.8''\times2.5''$ (with natural
weighting).

\begin{figure}
\includegraphics[angle=-90,width=0.45\textwidth]{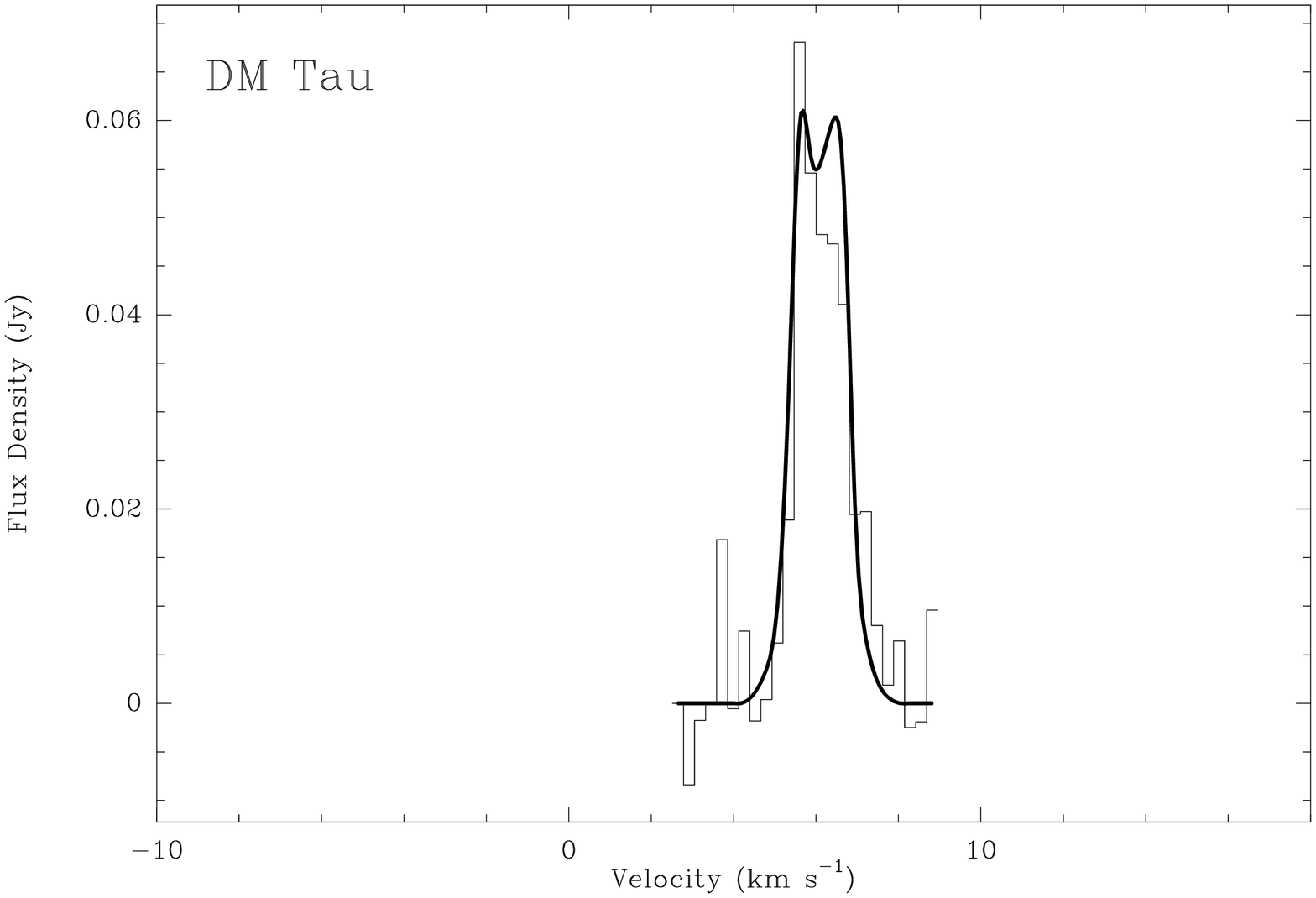}
\includegraphics[angle=-90,width=0.45\textwidth]{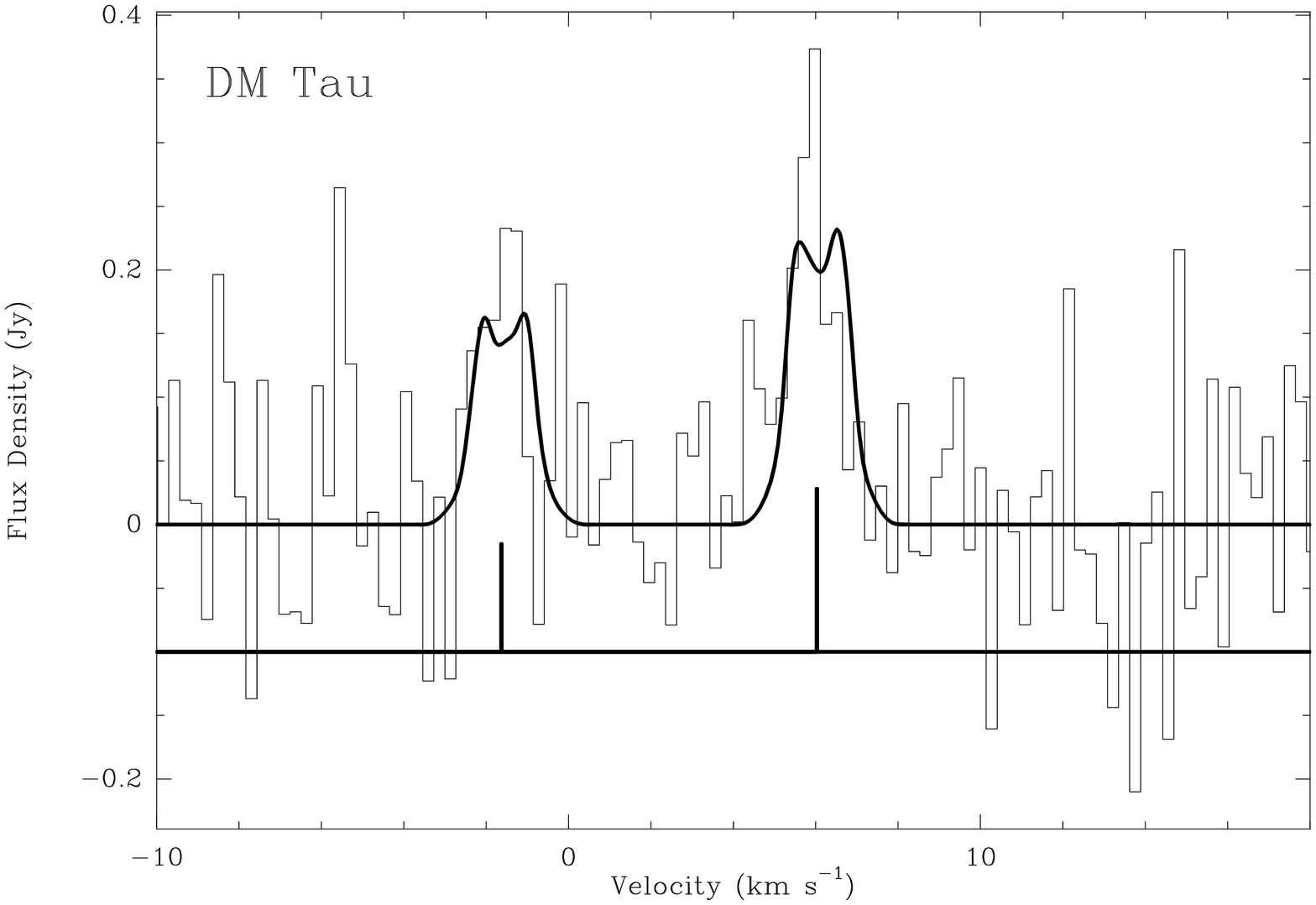}\\
\includegraphics[angle=-90,width=0.45\textwidth]{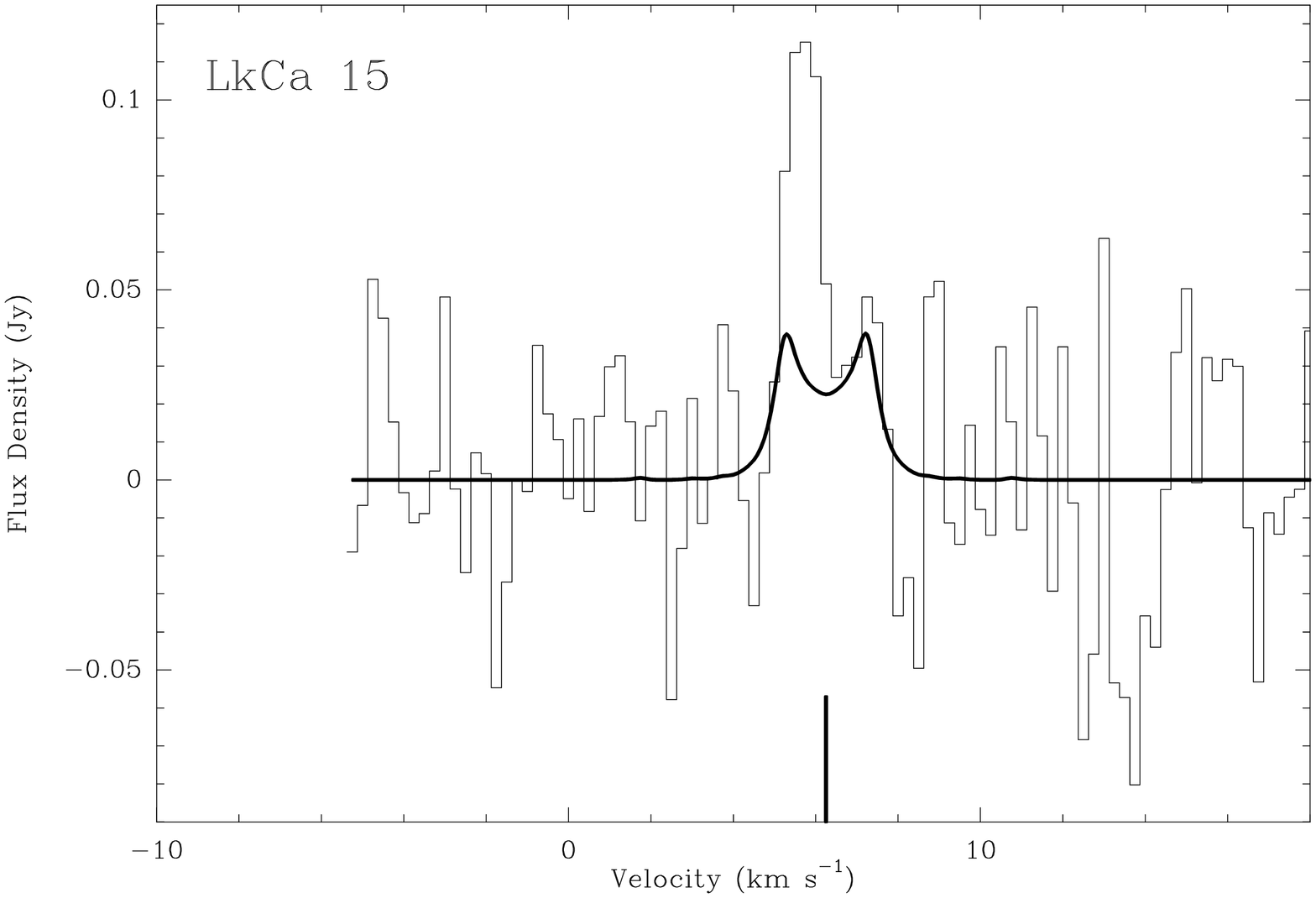}
\includegraphics[angle=-90,width=0.45\textwidth]{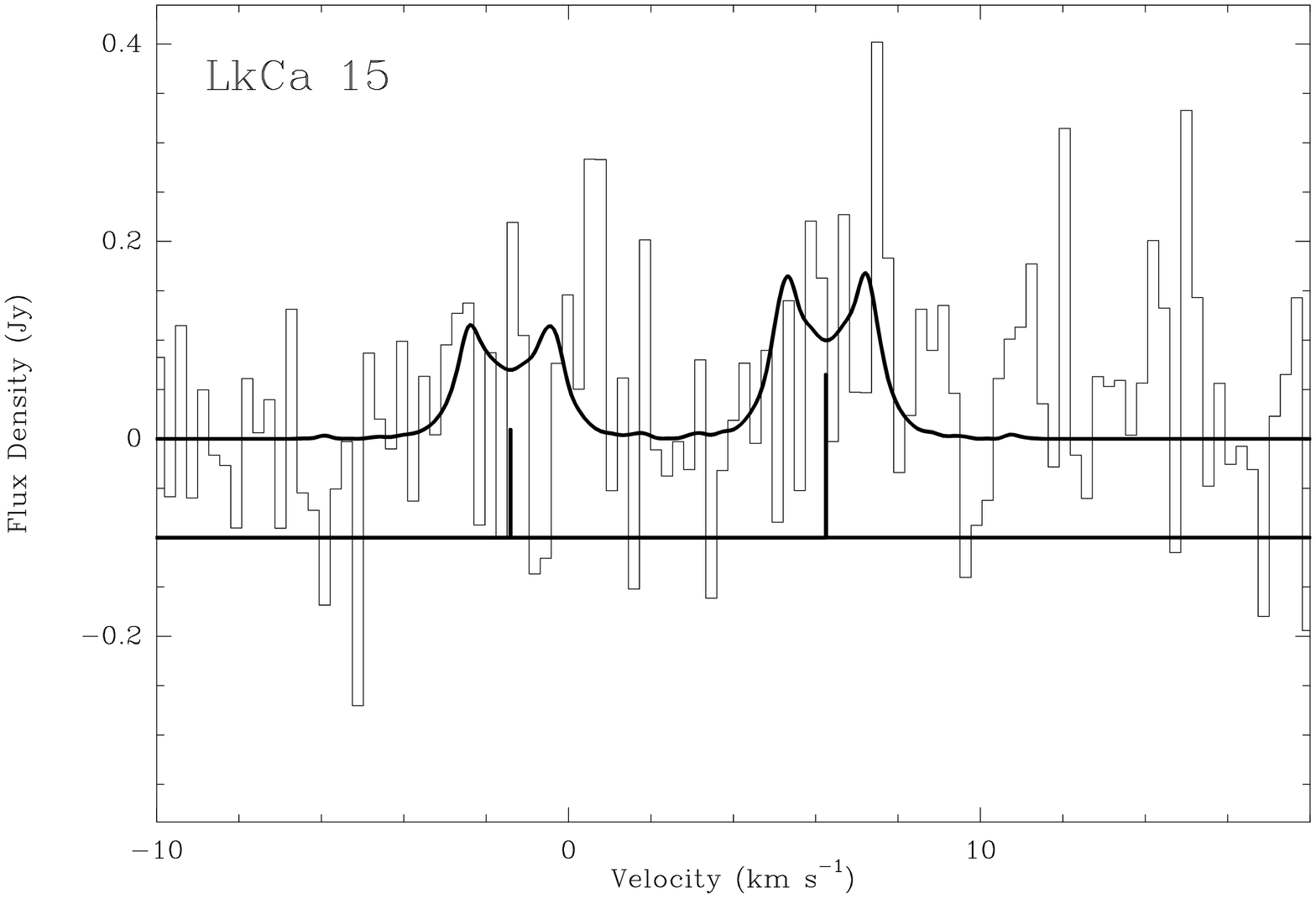}\\
\includegraphics[angle=-90,width=0.45\textwidth]{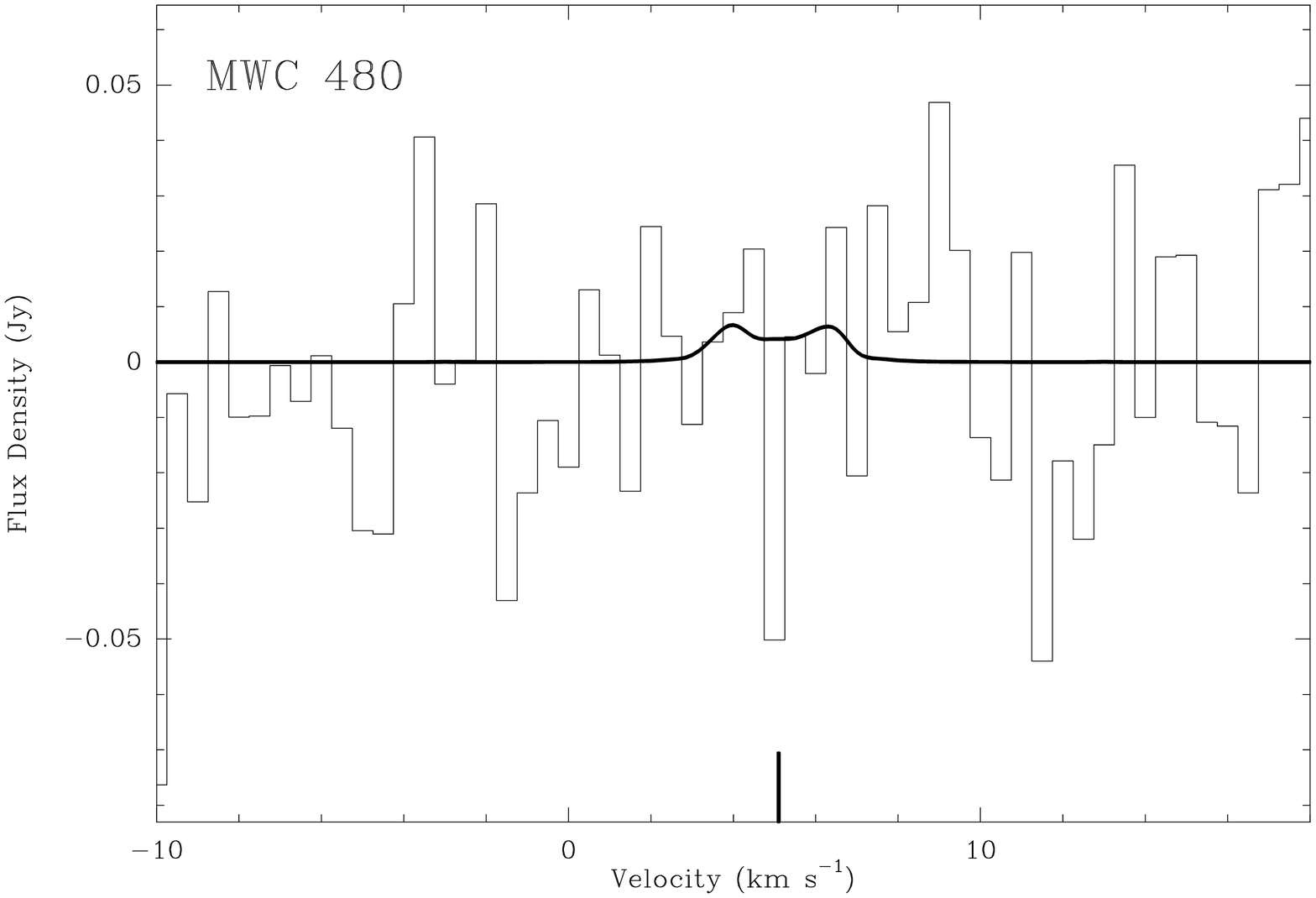}
\includegraphics[angle=-90,width=0.45\textwidth]{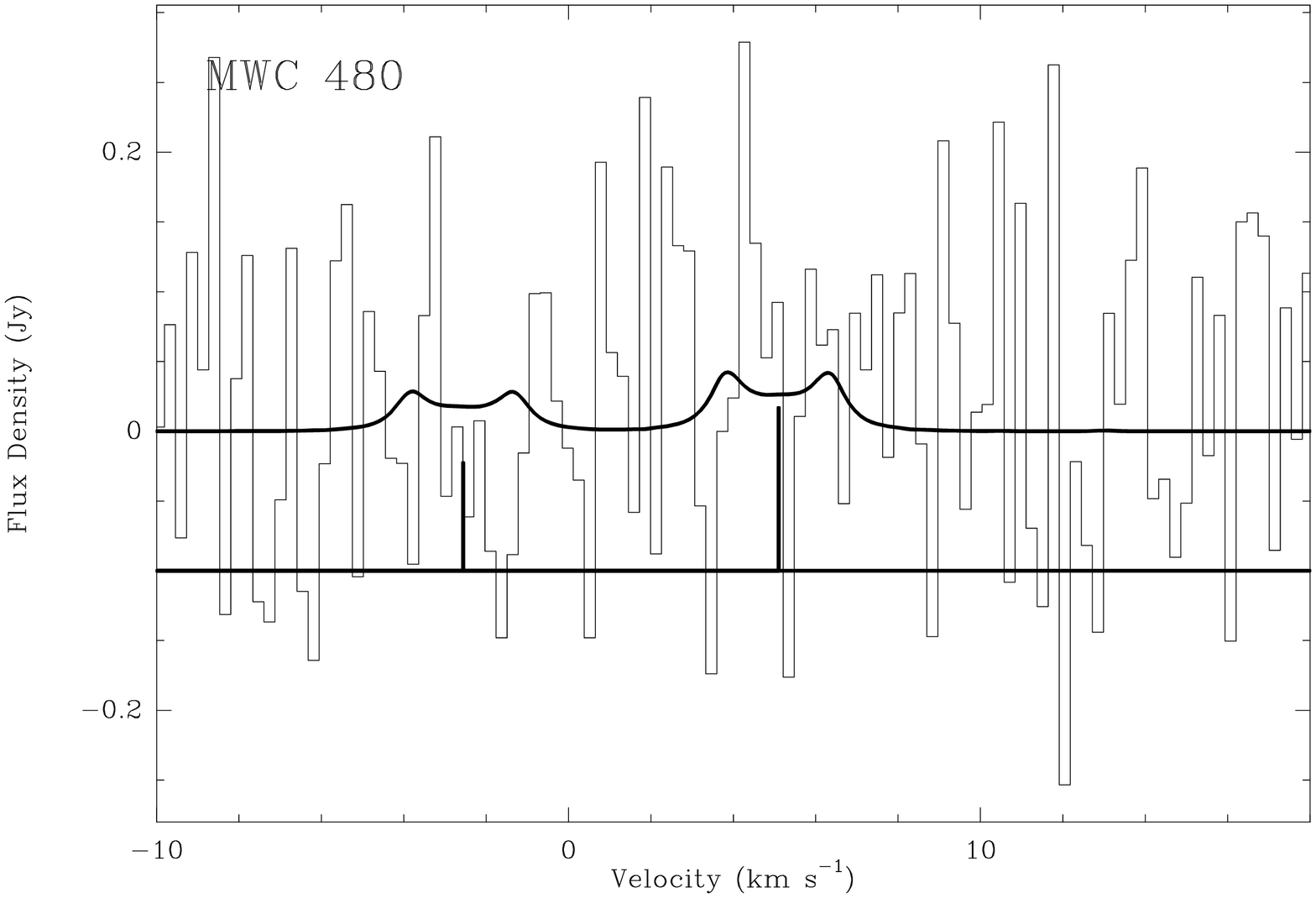}\\
\caption{Integrated spectrum of C$_2$H $J=1-0$ (left) and $J=2-1$ (right) for
DM~Tau, LkCa~15
and MWC~480 integrated over the $7''\times7''$ central region. The thick curves
represent the best-fit model deduced from the $\chi^2$ minimization procedure.}
\label{fig:spectra}
\end{figure}

Data were reduced with the GILDAS package. Flux calibration was performed
using MWC~349. The continuum was subtracted for the analysis. The channel width
is 0.25 km\,s$^{-1}$ for the J=1-0 line and 0.26 km\,s$^{-1}$ for the J=2-1
line. Integrated spectra over a $7\times 7''$ region (for DM Tau) and $5\times
5''$ region (for MWC~480 and LkCa~15) are presented in Fig.~\ref{fig:spectra}.
We detected C$_2$H 1-0 emission from DM\,Tau and LkCa\,15, but not from
MWC\,480.
The weak J=2-1 transition is detected in the T Tauri sources (see
Fig.~\ref{fig:spectra}). The apparent
disagreement between the observed
and modeled line profiles for the DM Tau and LkCa~15 spectra are due to the
relatively low signal-to-noise (SNR) level of our PdBI data.

\begin{figure}
\includegraphics[angle=-90,width=0.8\textwidth]{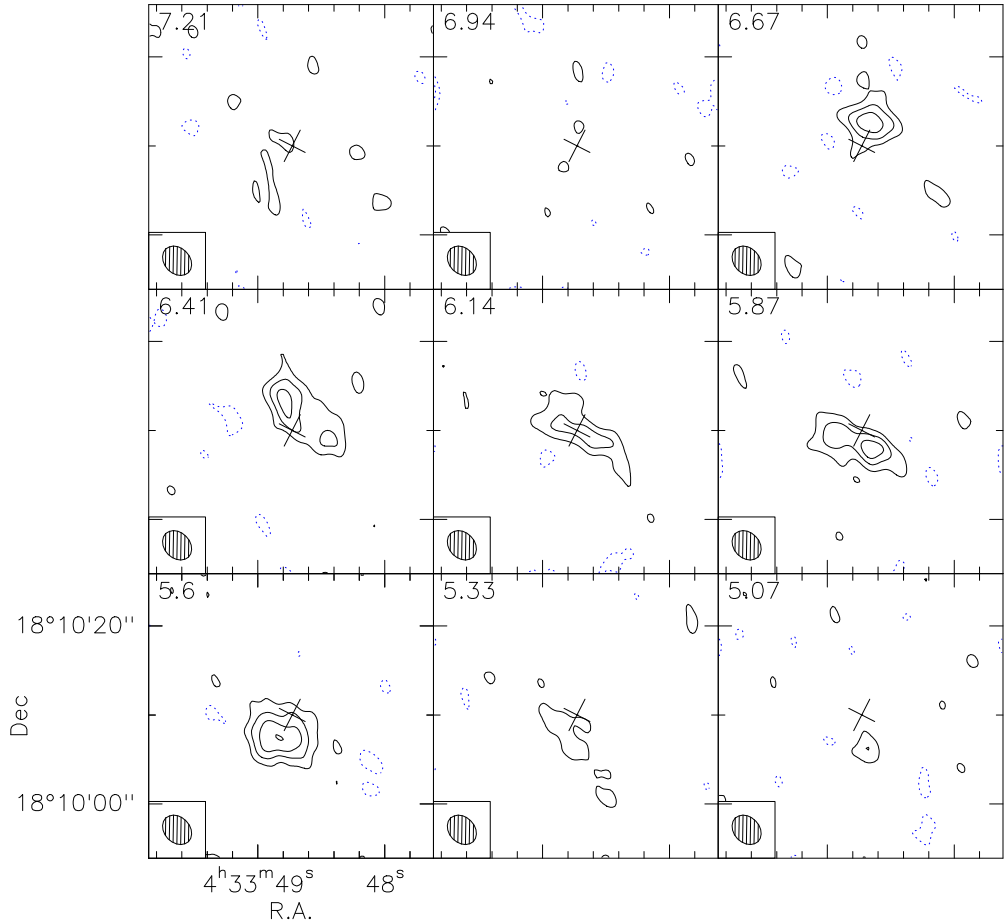}
\caption{Channel maps of the C$_2$H $J=1-0$ emission towards DM~Tau. Contour
spacing is 11~mJy/beam, or 0.17~K ($2 \sigma$). Velocities are given in the
upper left corners of the panels. The cross indicates the orientation of the
disk. The beam sizes are indicated in the lower left corners.}
\label{fig:maps}
\end{figure}

\begin{figure}
\includegraphics[width=0.4\textwidth]{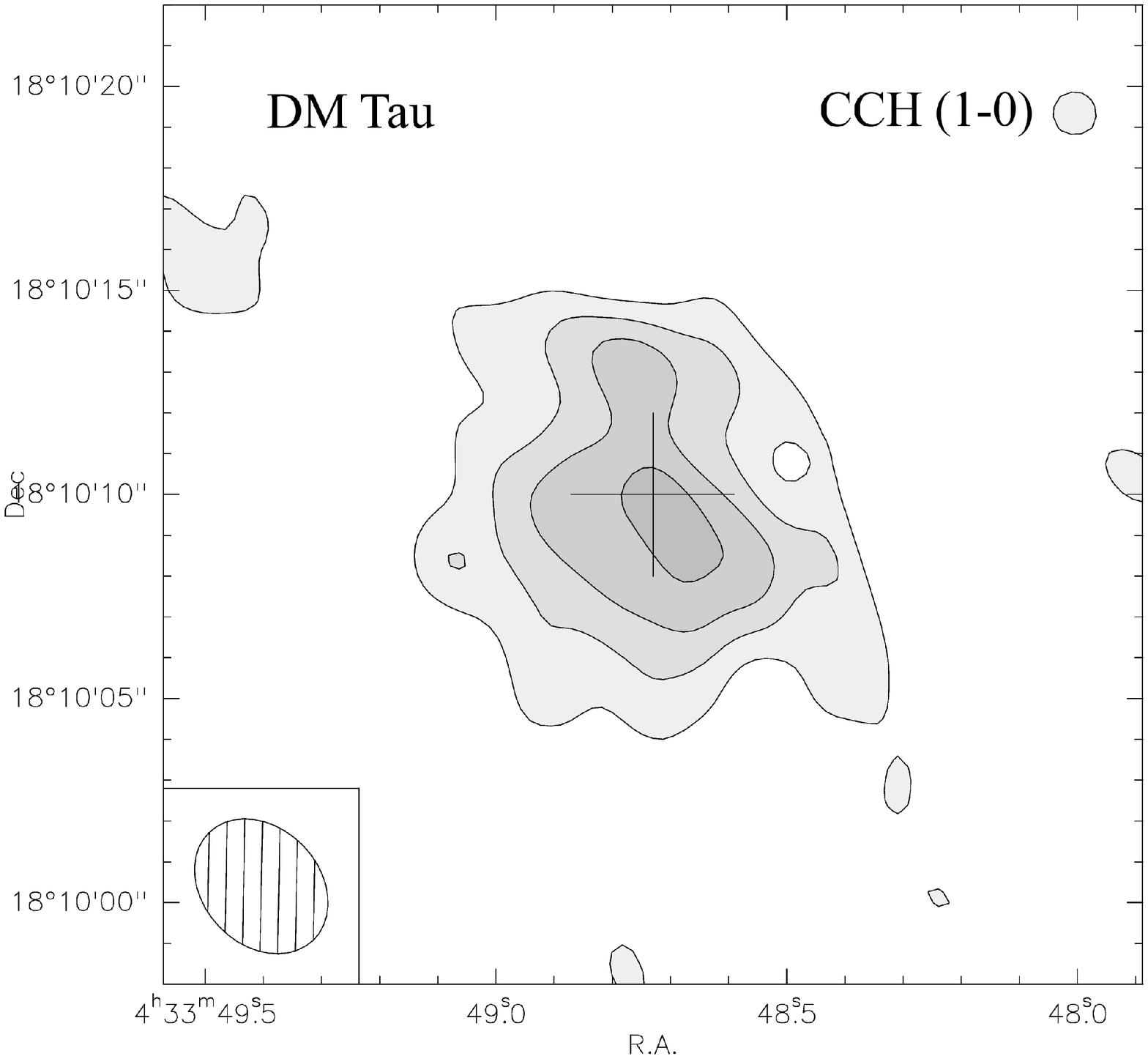}
\includegraphics[width=0.4\textwidth]{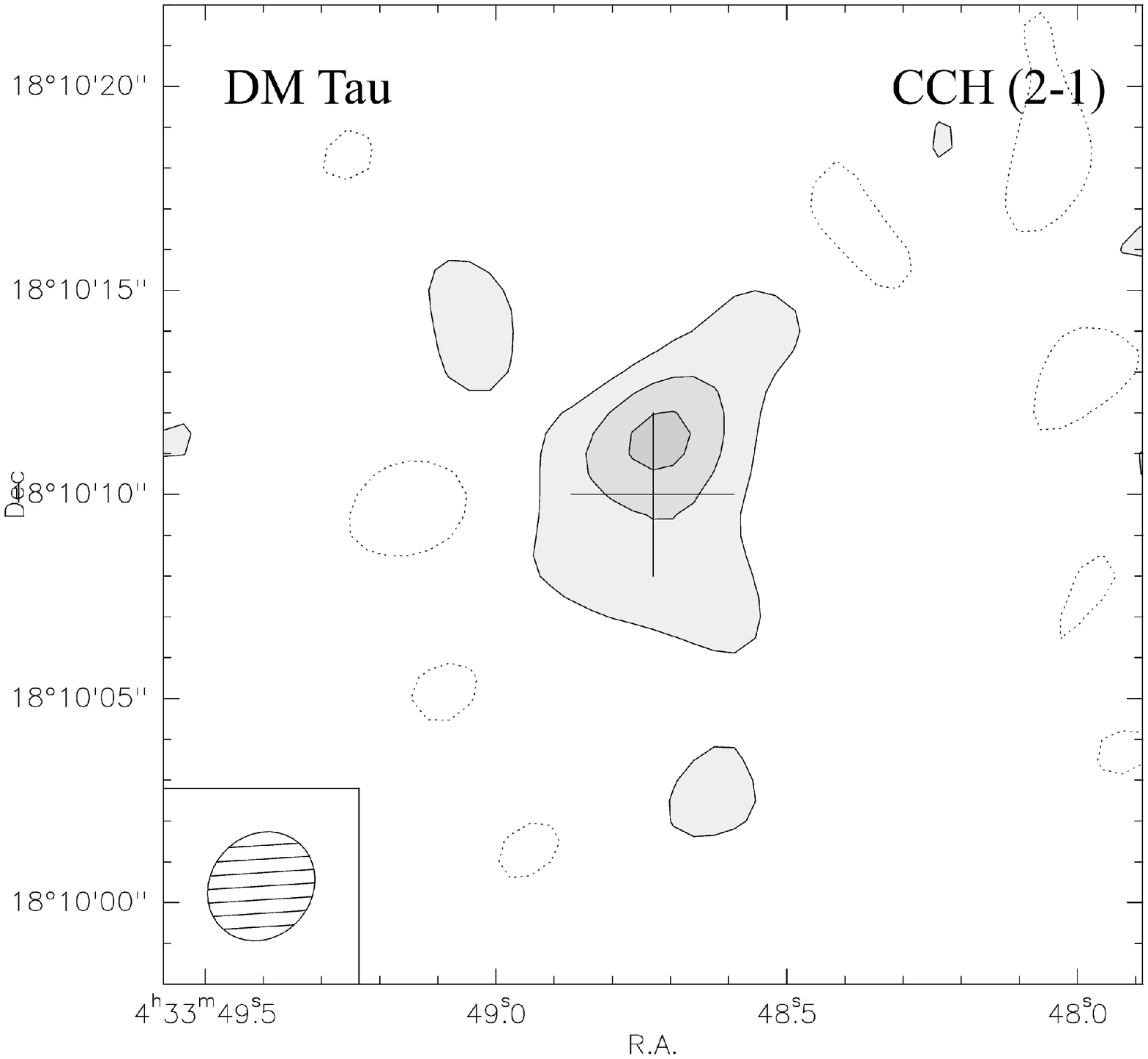}\\
\includegraphics[width=0.4\textwidth]{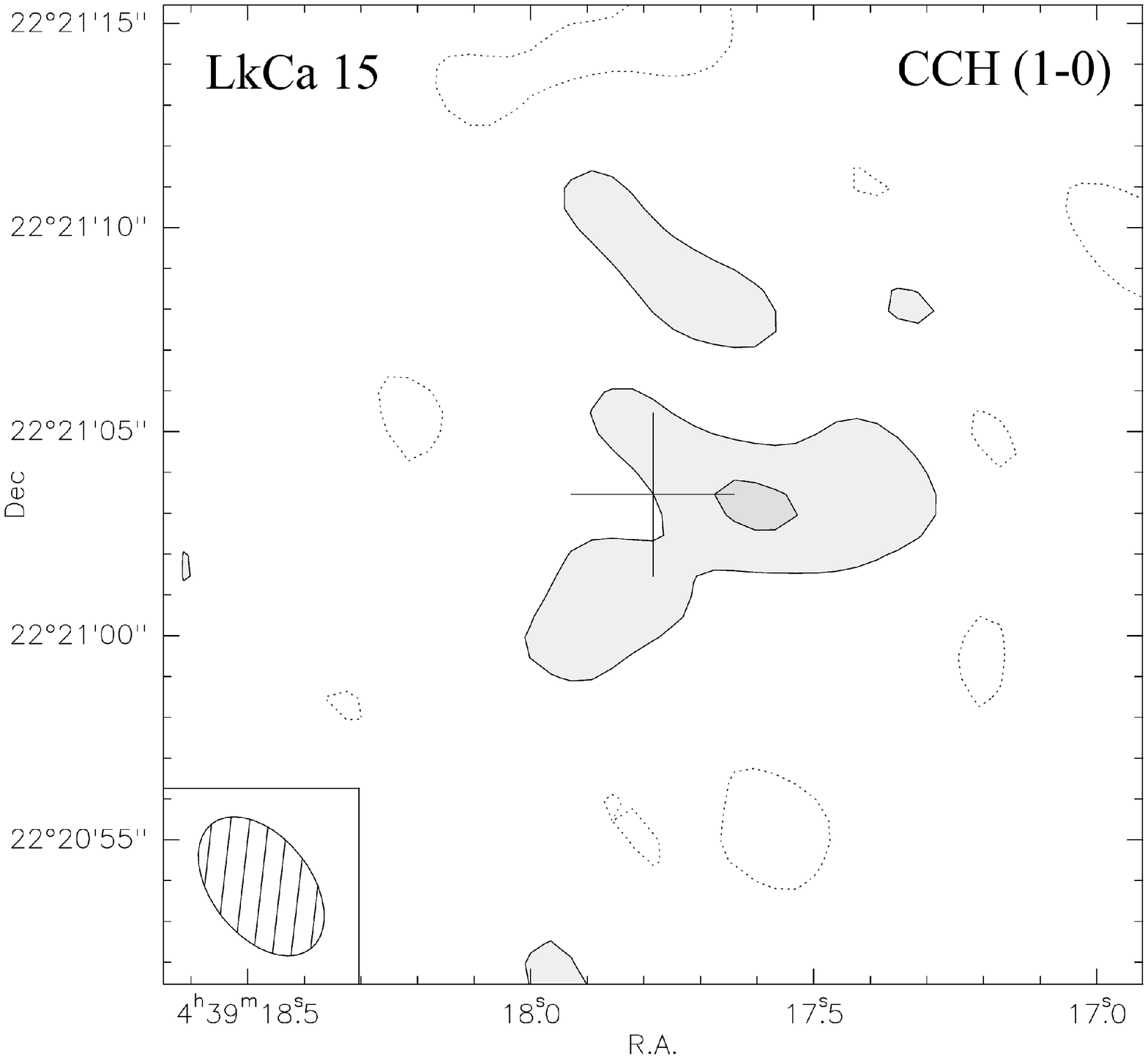}
\includegraphics[width=0.4\textwidth]{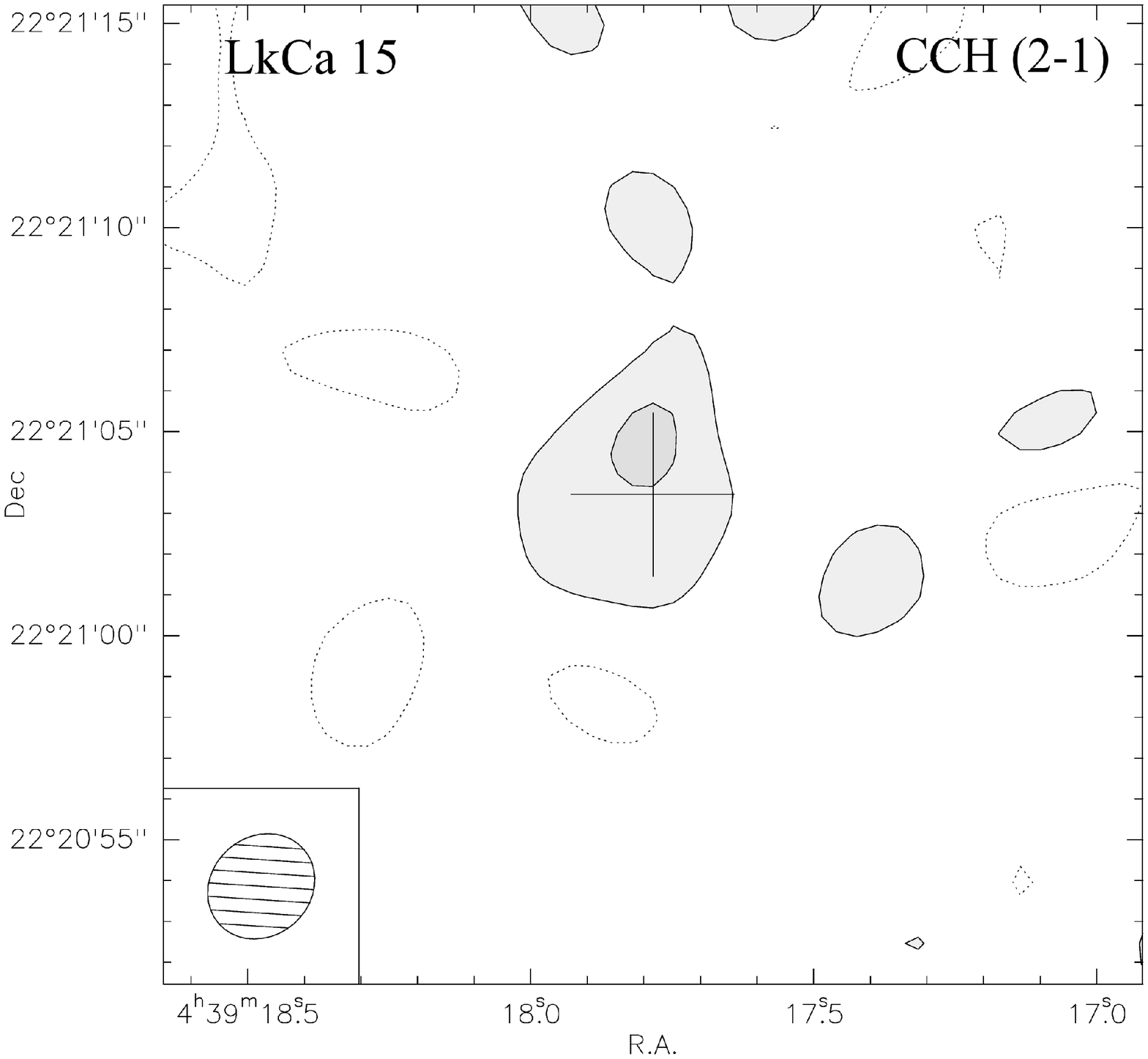}\\
\includegraphics[width=0.4\textwidth]{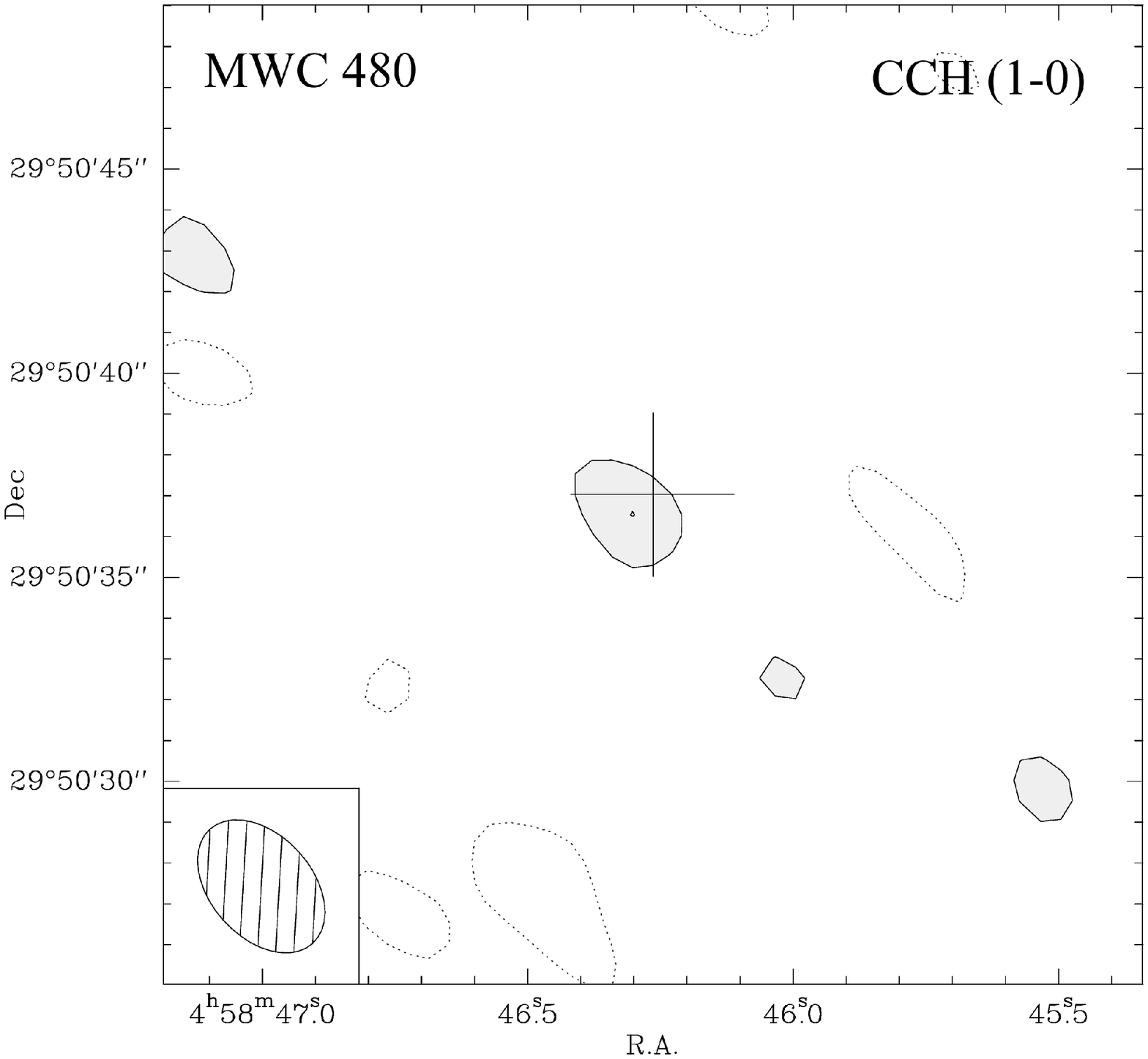}
\includegraphics[width=0.4\textwidth]{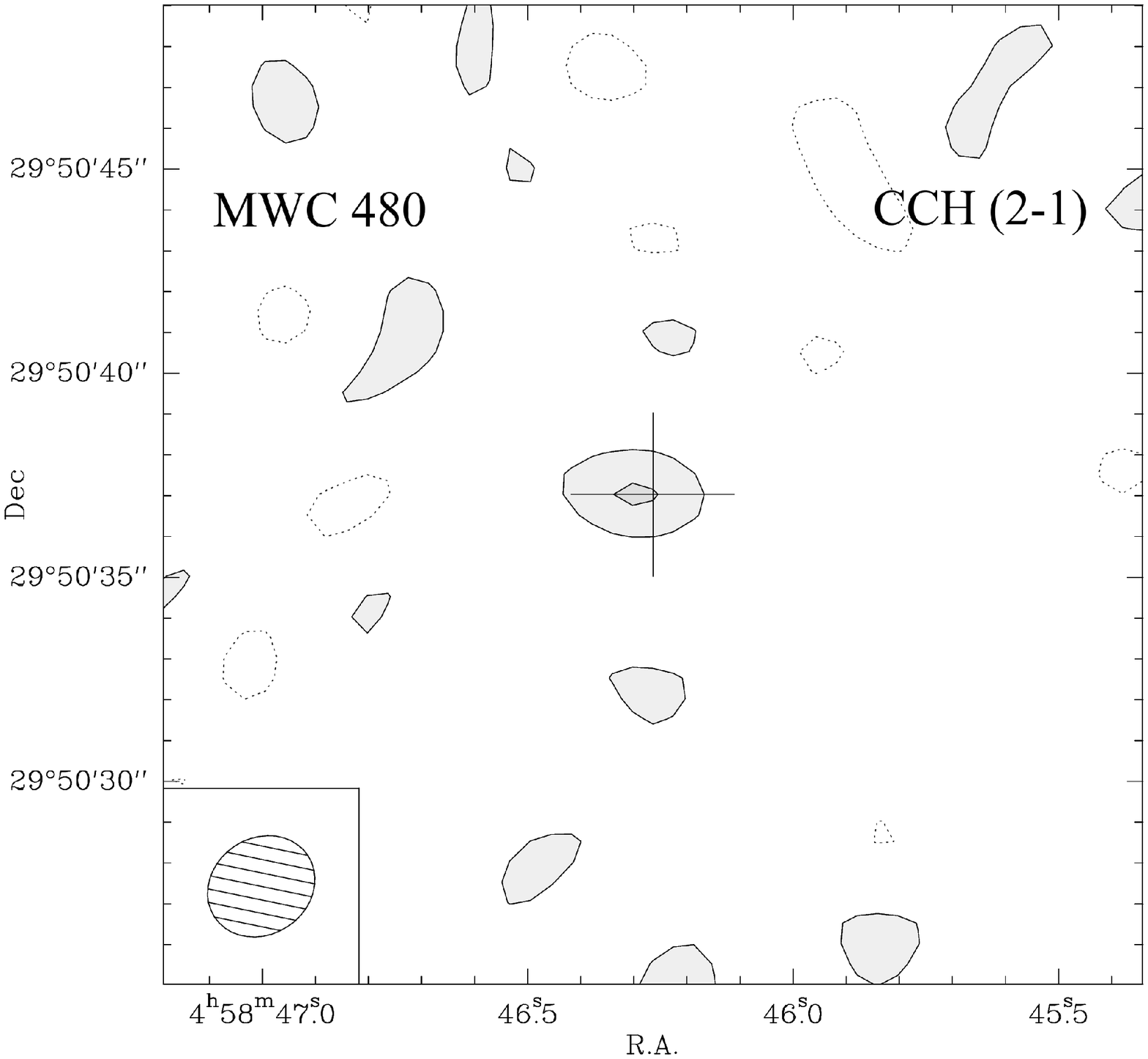}\\
\caption{(Top to bottom) SNR maps of the C$_2$H $J=1-0$ (left) and
$J=2-1$ (right) toward DM~Tau, LkCa~15 and MWC~480. The contours correspond to
the 2 $\sigma$ signal. The beam size is indicated in the lower left corners.}
\label{fig:sn}
\end{figure}

Channel maps of
the C$_2$H (1-0) emission in DM\,Tau are presented in Fig.~\ref{fig:maps}.
Since the signal-to-noise ratio of the C$_2$H (1-0) spectra from MWC~480
and LkCa~15 is low, we analyzed these data as in our study of N$_2$H$^+$ J=1-0
in the same sources \citep{Dutrey_ea07}. A filter to the spectra is applied, in
which each velocity channel is weighted by the expected line intensity (obtained
from the best fit modeling), and the sum over the channels is normalized by the
sum of the weights (a classical optimal filtering). The resulting SNR maps are
presented in Fig.~\ref{fig:sn}, which illustrate marginal
detection of C$_2$H (1-0) in LkCa~15 and MWC~480 \citep{Dutrey_ea07}.

\subsection{Derived column densities}
\label{obs_fit}
The distribution of C$_2$H in DM\,Tau, LkCa~15 and MWC\,480 was
analyzed using a parametric disk model, in which the column density
and excitation temperatures are power laws of the radius \citep[see][for a more
detailed description of the model]{Pietu_ea07}. We used the geometric
parameters
(position, inclination, Keplerian velocity law) derived from the CO
isotopologues and HCO$^+$ (see Table~\ref{tab:disk_struc}). The
$\chi^2$-minimization
technique of \citet{GD98} was applied to constrain the CCH column densities
and other disk parameters.
The observation of two  C$_2$H lines allows us to recover
the excitation conditions (within the limitations due to low signal to noise
data). As the lines are relatively weak, the exponent of the excitation
temperature is not well constrained. We find $q =0.4 \pm 0.3$ for DM\,Tau, and
assumed $q=0.4$ for the other sources. Such a value is in agreement with results
obtained from stronger lines, in particular HCO$^+$
(Table~\ref{tab:disk_struc}). Consequently, the excitation temperature of the
DM\,Tau disk is well constrained, and poorly derived for LkCa~15. For MWC\,480
we had to assume $T = 14$~K in the data analysis.
The obtained column densities are insensitive to the choice of the
temperature in a range between 10 and $\sim 20$~K at 100~AU, and scale linearly
with T for
higher temperatures. The slope of the surface density distribution is
constrained for DM\,Tau and Lk\,Ca\,15. For MWC\,480 we used a similar value.
The derived temperature and column densities are summarized in
Table~\ref{tab:c2hcd}.

\begin{deluxetable}{crrr}
\tablecaption{Derived column densities of CCH
\label{tab:c2hcd}}
\tablehead{\colhead{Source} & \colhead{DM Tau} & \colhead{LkCa 15} &
\colhead{MWC 480}}
\startdata
$\Sigma$ at 300 AU ($10^{13}$ cm$^{-2}$) & $2.8 \pm 0.2$ & $ 2.9 \pm 1.1$
& $\le 0.8\pm0.3$ \\
$p$ & $0.5\pm 0.2$ & $0.5 \pm 0.2$ & [0.5] \\
$R_\mathrm{out}$ (AU) & $780 \pm 40$ & $570 \pm 50$ & [400] \\
$T$ at 300 AU (K) & $7.3\pm0.6$ & $6.3\pm 1.4$ & [9] \\
$q$ & $0.4\pm 0.3$ & [0.4] & [0.4] \\ \enddata
\tablecomments{Values between square brackets indicate fixed parameters
as constrained by the previous CO, HCO$^+$ and new CCH data fitting.
\citep{Pietu_ea07}. The derived CCH column density for MWC 480 is an upper
limit.
}
\end{deluxetable}

\section{Disk physical and chemical model}
\label{model}
We refrain from comparing
deconvolved
observed and simulated spectral maps and base our analysis on the
observed quantities derived with the $\chi^2$-minimization method in
the {\it uv}-plane \citep{GD98}. This allows us to exclude computationally
expensive iterative fitting of the data using 2D line radiative
transfer modeling.

\begin{figure}
\includegraphics[angle=90,width=0.65\textwidth]{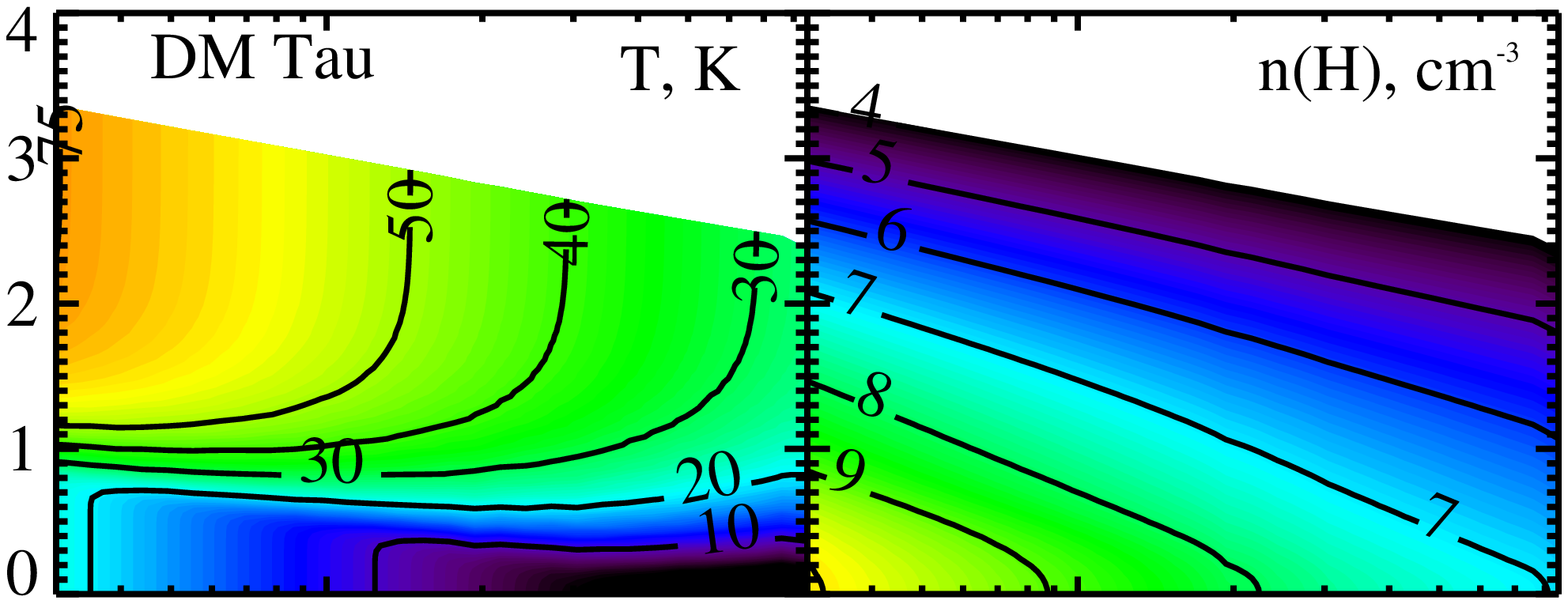}\\
\includegraphics[angle=90,width=0.65\textwidth]{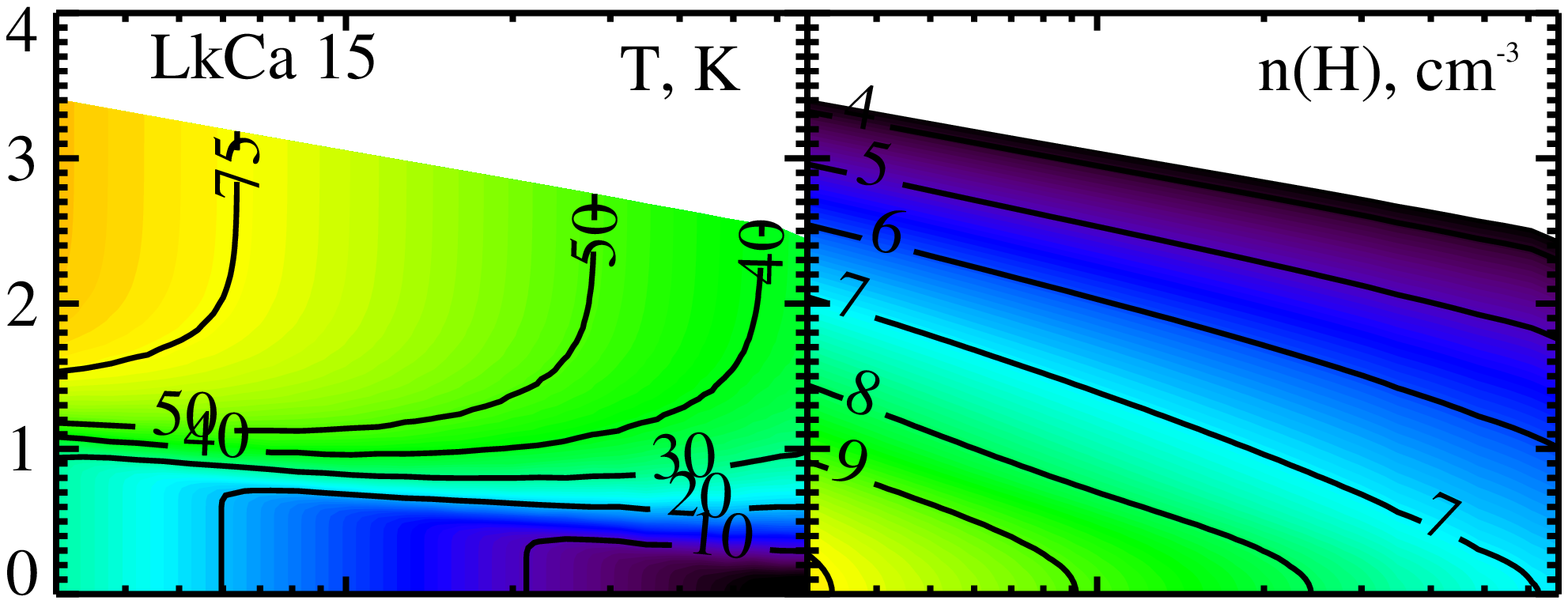}\\
\includegraphics[angle=90,width=0.65\textwidth]{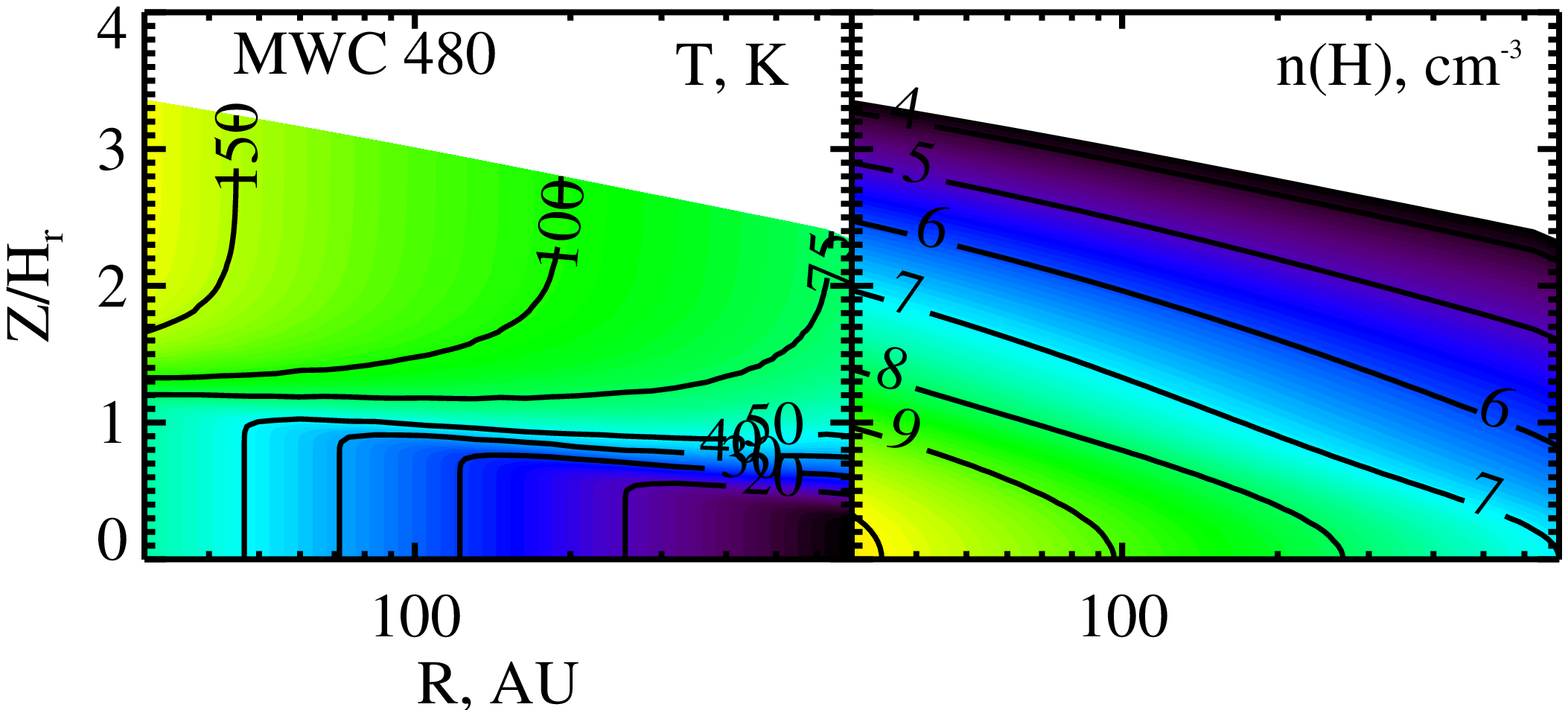}
\caption{Distributions of the dust temperature (left panel) and particle
density (right panel, logarithmic scale) for the DM~Tau, LkCa~15, and MWC~480
disk models (top to bottom). The gas and dust temperatures are assumed to be
the same. The Y-axis is given in units of the pressure scale height (as derived
for the disk atmosphere).}
\label{fig:disk_struc}
\end{figure}

For all sources, we adopted a flared 1+1D-disk model with a vertical
temperature gradient similar to that of \citet{DAlessio_ea99}, see
Fig.~\ref{fig:disk_struc}.
The $\alpha$-parameter for all three disks is chosen such that the
modeled disk masses are in accordance with the observed values. Input disk and
stellar parameters are either determined by the
$\chi^2$-fitting or taken from previous works
\citep{Dutrey_ea97,Mannings_ea97,DAlessio_ea99,Simon_ea00,Thi_ea04,Calvet_ea05,
Pietu_ea07}, see Table~\ref{tab:disk_struc}.
The adopted viscosity parameter $\alpha$ is 0.01 for DM Tau, 0.03 for LkCa~15,
and 0.02 for MWC~480. It is chosen such that the resulting disk masses are
the same as derived from millimeter dust continuum measurements.
The dust grains are modeled as compact spheres
of uniform 0.12~$\mu$m radius made of amorphous silicates with olivine
composition and the optical data taken from \citet{DL84}.  The standard $1\%$
dust-to-gas mass ratio is assumed. In addition, we consider a model of the DM
Tau disk with bigger $1.0\mu$m grains.

The disks are illuminated by the UV radiation from the central star
and by the interstellar UV radiation. The intensity of the interstellar UV
radiation field is $\chi=1$ as derived by \citet{G}. The UV intensity at
a given disk
location is calculated as a sum of the stellar and interstellar
components that are scaled down by the visual extinction in vertical
direction and in direction to the central star (1D plane-parallel
approximation). The non-thermal FUV radiation field from DM Tau and LkCa~15 is
represented by the scaled ISRF, while that of MWC~480 is represented by
the scaled radiation field of the Herbig A0 star (according to
\citet{vD88} and \citet{vDea_06}).
We model the attenuation of cosmic rays (CRP) by
Eq.~(3) from \citet{Red2}.  In the disk interior, ionization due to
the decay of short-living radionuclides is taken into account,
assuming an ionization rate of $6.5\cdot10^{-19}$~s$^{-1}$
\citep{Finocchi_Gail97}.  The intensity and radial penetration of the stellar
X-ray radiation are modeled using observational results of
\citet{Glassgold_ea05} and the 2D Monte Carlo simulations of
\citet{zetaxa,zetaxb}.

We did not find any relevant data regarding observed X-ray luminosities
for our stars and used the following estimates: $L_{\rm X} =
10^{28}$~erg\,s$^{-1}$ for MWC~480, and $3\,10^{29}$~erg\,s$^{-1}$ for DM~Tau
and for LkCa~15. The low value of the adopted X-ray
luminosity of MWC 480
reflects the fact that Herbig Ae stars have weak surface magnetic fields due
to their non-convective interiors and thus lack an $\alpha\omega$ dynamo
generation mechanism. Consequently, their coronal activity is lower and X-ray
luminosities are 10-1000 times lower than that of T Tauri stars
\citep{Guedel_Naze09}. Indeed, a few measurements of the strength of
the longitudinal magnetic field in MWC 480 resulted in
marginal detection of $B = 87 \pm 22$~Gauss \citep{Hubrig_ea04,Hubrig_ea06}.
For the T Tauri sources we adopted a median value as recently measured with
Chandra and XMM in the Orion Nebula Cluster and the Cepheus B
star-forming region \citep{Preibisch_ea05,Getman_ea09}: $\log (L_{\rm X}/L_{\rm
bol}) \approx -3.5 = 3\,10^{29}$~erg\,s$^{-1}$ (with uncertainty of an order of
magnitude).
To investigate the role of the stellar X-ray
radiation for the CCH chemical evolution, we consider three additional
models of the DM Tau disk with $L_{\rm X}=0, 10^{29}$ and
$10^{30}$~erg\,s$^{-1}$.

We added a new class of the X-ray driven reactions to our chemical code
leading to the formation of doubly-charged atoms (O$^{++}$, C$^{++}$, N$^{++}$,
S$^{++}$, Fe$^{++}$, Si$^{++}$)
due to the Auger effect. Secondary electron impact ionization cross sections
were taken from \citet{MS_05}. Neutralization reactions of these
doubly-charged ionized atoms by e- and charge transfer reactions with molecules
are adopted from \citet{Staeuber_ea05}. The photoionization cross sections are
taken from \citet{Verner_ea93}, as described in \citet{Maloney_ea96}. In
addition, we duplicated a few tens reactions of CRP with molecules to get the
same reaction set for the X-ray photons. The rate of CO was modified in order to
take self-shielding properly into account according to the Eq.~9
from \citet{Maloney_ea96}.

The gas-grain time-dependent chemical model
adopted in this study is mostly the same as in the CID2 paper.
First, the osu.2007 network of gas-phase reactions\footnote{See: {\it
https://www.physics.ohio-state.edu/$\sim$eric/research.html}} is used.
Second, we utilize a standard rate approach to the surface chemistry modeling
but without H and H$_2$ tunneling \citep{Katz_ea99}.
The surface reactions together with desorption energies were adopted from
the model of \citet{Garrod_Herbst06}, which is mostly based on previous studies
of \citet{Hasegawa_ea92} and \citet{Hasegawa_Herbst93}.
Finally, rates of several tens of photodissociation and photoionization
reactions are updated according to \citet{vDea_06}, using the UV spectral shape
typical of a Herbig Ae star and ISM-like
dust grain optical properties. Overall, the disk chemical network consists of
more than 650 species made of 13 elements and 7500 reactions.

Using this time-dependent model and the ``low metals'' initial
abundances of \citet{Lee_ea98}, distributions of the molecular abundances
and column densities for the considered species are simulated over 5~Myr of
evolution. The choice of initial abundances does not affect the
resulting molecular abundances and column densities due to relatively high
densities in and long evolutionary timescales of protoplanetary disks
\citep[see also][]{Willacy_ea06}.

\section{Results of chemical modeling}
\label{results}

In Figure~\ref{fig:cch_abunds}, the radial and vertical distribution
of the absolute and relative C$_2$H abundances (to the total amount of
hydrogen) at 5~Myr are shown. The highest relative C$_2$H abundance is only
about $10^{-9}$ (in DM~Tau and LkCa~15) and several times lower in MWC~480. In
DM~Tau and LkCa~15 the warm layer of ethynyl is located at $\sim 1$ pressure
scale height, with a local peak towards $r=10$~AU. Note that there is a second
peak, better visible in relative concentration, which is located in the outer
disk region and at higher disk heights ($r \ga 500$~AU, $z/H_r \sim 1.5$), where
the IS UV photons can penetrate. CCH is partly frozen out in all T~Tau disk
midplanes as its evaporation temperature is $\sim 50$~K.
In the warmer disk of MWC~480 ethynyl reaches the highest relative
abundance only in the outer disk region, at $H_r \sim 1$, while a peak in
absolute CCH concentration is reached in the midplane around 100~AU.
What is the reason for this chemical diversity of CCH in all 3 disks?

\begin{figure}
\includegraphics[width=0.95\textwidth]{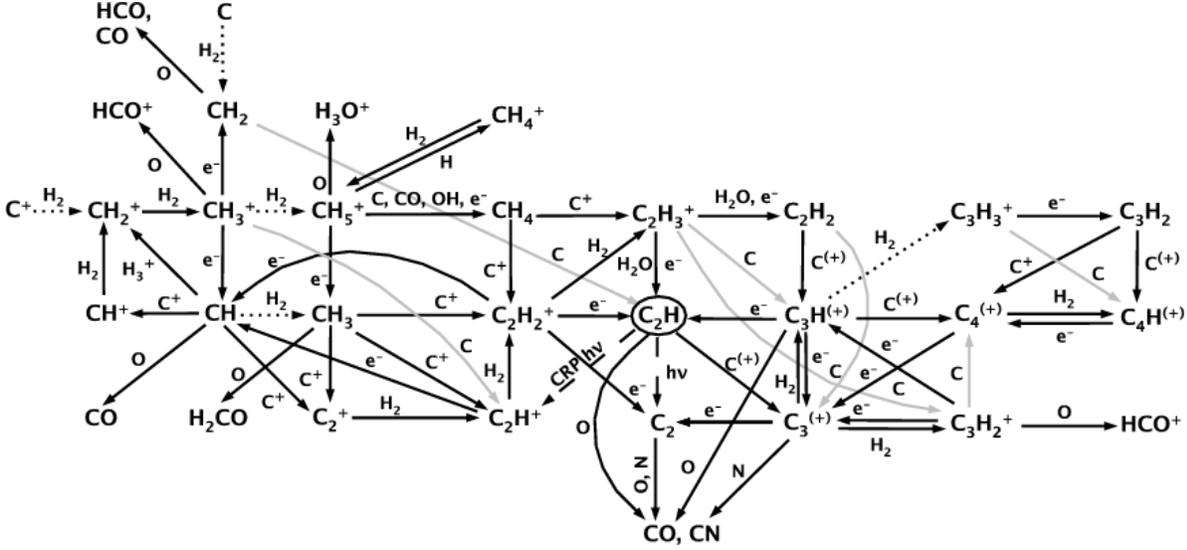}
\caption{Most important chemical pathways for the evolution of CCH in
  protoplanetary disks.}
\label{fig:cch_network}
\end{figure}

In essence, the chemical evolution of the ethynyl radical is determined by
removal rates of carbon and oxygen into more complex molecules,
e.g. CO, OH, H$_2$O, as well as by photoprocesses and freeze-out/evaporation,
see Fig.~\ref{fig:cch_network}. Our chemical simulations begin with
atomic initial abundances.  Thus, light ionized
hydrocarbons CH$^+_{\rm n}$ (n=2..5) are the first molecular species produced by
radiative association with H$_2$, following hydrogen addition reactions: C$^+$
$\rightarrow$ CH$_2^+$ $\rightarrow$ CH$_3^+$ $\rightarrow$ CH$_5^+$.  The
protonated methane reacts with electrons, CO, C, OH, and more complex species at
later stage and forms methane. The CH$_4$ molecules undergo reactive
collisions with C$^+$, producing C$_2$H$_2^+$ and C$_2$H$_3^+$. An
alternative way to produce C$_2$H$_2^+$ is the dissociative
recombination of CH$_5^+$ into CH$_3$ followed by reactions with
C$^+$. The C$_2$H$_3^+$ ion is also formed by hydrogen addition to
C$_2$H$_2^+$.  Finally, C$_2$H$_2^+$ and C$_2$H$_3^+$ dissociatively
recombine into CH, C$_2$H, and C$_2$H$_2$. Another important formation channel
for ethynyl is by surface hydrogenation of precursor species, C$_2$.

The major removal pathway for C$_2$H is either the direct
neutral-neutral reaction with O that forms CO, or the same reaction
but with heavier carbon chain ions that are formed from ethynyl by
subsequent insertion of carbon. In upper disk regions photodissociation of CCH
into C$_2$ and H (with the rate constant $k=5.1\,10^{-10}$~s$^{-1}$ and
A$_{\rm V}$ exponent $\gamma=1.9$) and photoionization
($k=10^{-10}$~s$^{-1}$, $\gamma=2.0$) are efficient \citep{vD88}.
At later times, in colder disk regions depletion and
gas-phase reactions with more complex species may enter into this
cycle. One can restore high concentrations of CCH obtained at
early times, $t\la 10-100$~years by destroying CO and enriching the gas with
elemental carbon. In disk it happens in the inner region either by reactions of
CO with ionized helium atoms produced by the stellar X-ray radiation, or
directly upon irradiation of CO by UV photons. This second rise in the CCH
concentration is typically reached only at very late times in our models, $t
\ga 1$~Myr.

The relative importance of X-ray and UV radiation is shown in
Fig.~\ref{fig:cch_abunds_fact}, using the disk around DM Tau as an example. The
prominent intermediate layer of CCH almost disappears in the model without X-ray
irradiation, and when grains grow beyond $\sim 1\mu$m. The lower opacities in
the model with ``big'' grains shifts the upper outer layer of CCH towards
midplane, following the enhanced penetration of the dissociating interstellar UV
radiation. Thus CCH is indeed a sensitive tracer of the UV radiation field.
The lack of X-ray irradiation results in a less
``sharp'' inner shape of the upper CCH layer (compare left and right panels in
Fig.~\ref{fig:cch_abunds_fact}), and low CCH abundances in the
intermediate layer. In the absence of X-ray radiation, ionized helium is only
produced by interactions with cosmic ray particles, and thus replenishment of
C$^+$ and O into the gas upon destruction of CO by He$^+$ is lower. Still,
there is an ``island'' of high CCH concentration in the intermediate layer at
$r\sim 200-300$~AU in the model of DM~Tau without X-rays.

MWC~480, as a Herbig Ae star, emits $\sim 3$ and $\sim 15$ times more UV photons
than LkCa~15 and DM~Tau, respectively, but has a weak X-ray luminosity (see
Table~\ref{tab:coord}). Consequently, the upper CCH layer is more noticeable
in the MWC~480 disk, where a larger fraction of elemental carbon can be kept
from being locked in CO and other species by photodissociation. Note that the
molecular-poor atmosphere is slightly more extended in the MWC~480 disk compared
to those of LkCa~15 and DM~Tau (upper panel in Fig.~\ref{fig:cch_abunds}). The
low X-ray luminosity of MWC~480 assumed in our modeling also leads to lower
concentration of CCH in the inner warm molecular layer ($r\la 50$~AU,
$\approx 1$ pressure scale height). However, there is a region around
the midplane
($50-200$~AU, up to $z/H_r\approx 1$) where absolute abundances of CCH are high
in the MWC~480 disk. Since the Herbig Ae disk is warm, surface processes
involving heavy radicals are more active there, while volatile ices are nearly
absent. Using our chemical analyzing software, we found that the oasis of CCH
in MWC~480 is due to efficient surface hydrogenation of C$_2$ ice, and other
surface reactions (e.g., photodissociation of C$_2$H$_2$ ice, and surface
formation of C$_2$ from two C atoms). Temperature there ranges from 23--50~K,
while densities are high, $n_{\rm H} \approx 10^6-10^8$~cm$^{-1}$, resulting in
fast accretion rates of precursor molecules to grain surfaces and sufficiently
high desorption rate of CCH. Note that this oasis of ethynyl appears only after
$\approx 1$~Myr of evolution, due to slow surface chemistry.
In both T Tau disks photodissociation of CO is effective in the outer
upper disk region. Both DM~Tau and LkCa~15 have non-negligible X-ray luminosity
of $\sim 10^{30}$~erg\,s$^{-1}$ \citep{Glassgold_ea05}, and thus CCH can be
re-formed toward inner disk regions, see
Fig.~\ref{fig:cch_abunds}.

\begin{figure}
\includegraphics[angle=90,width=0.31\textwidth]{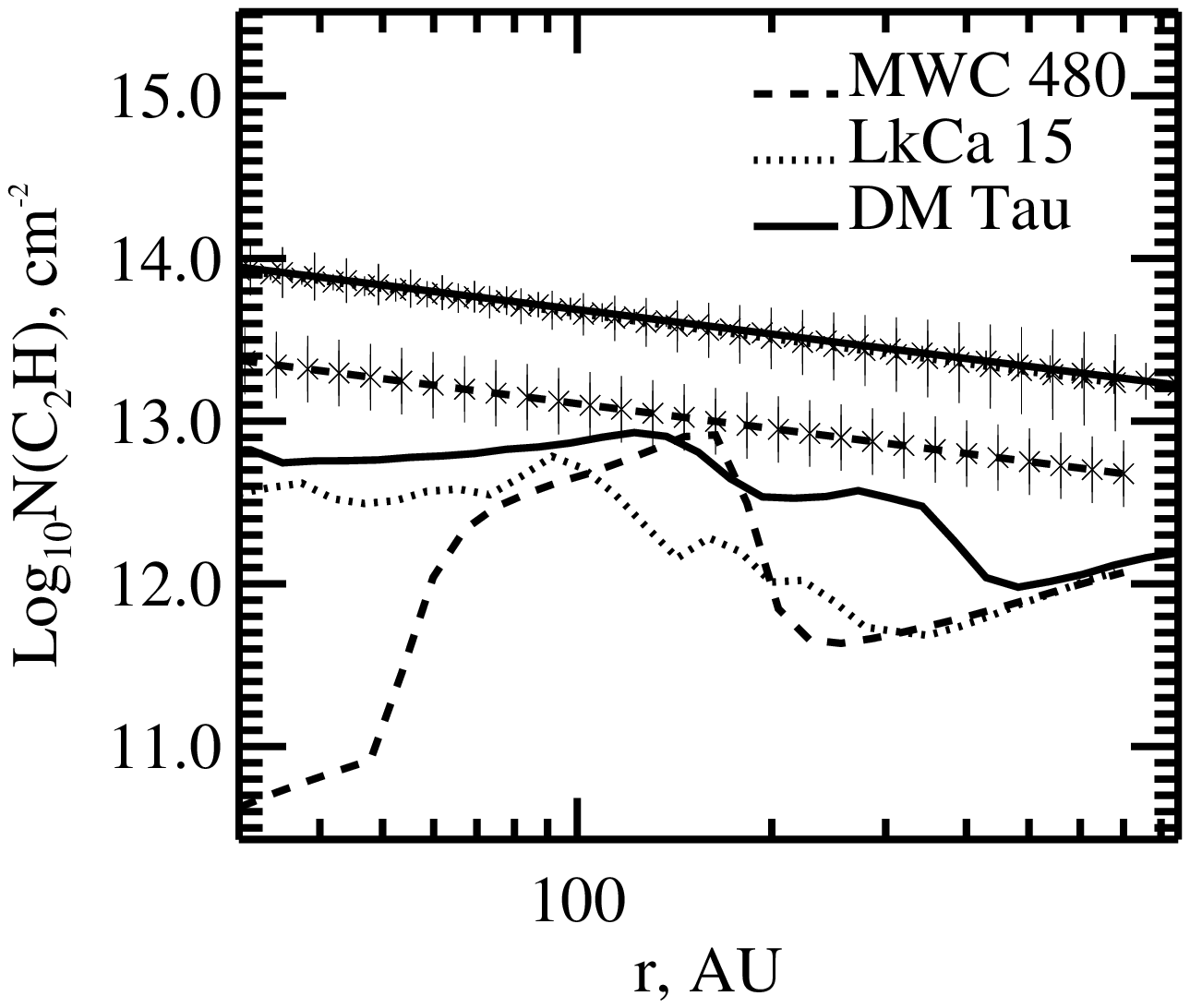}
\includegraphics[angle=90,width=0.31\textwidth]{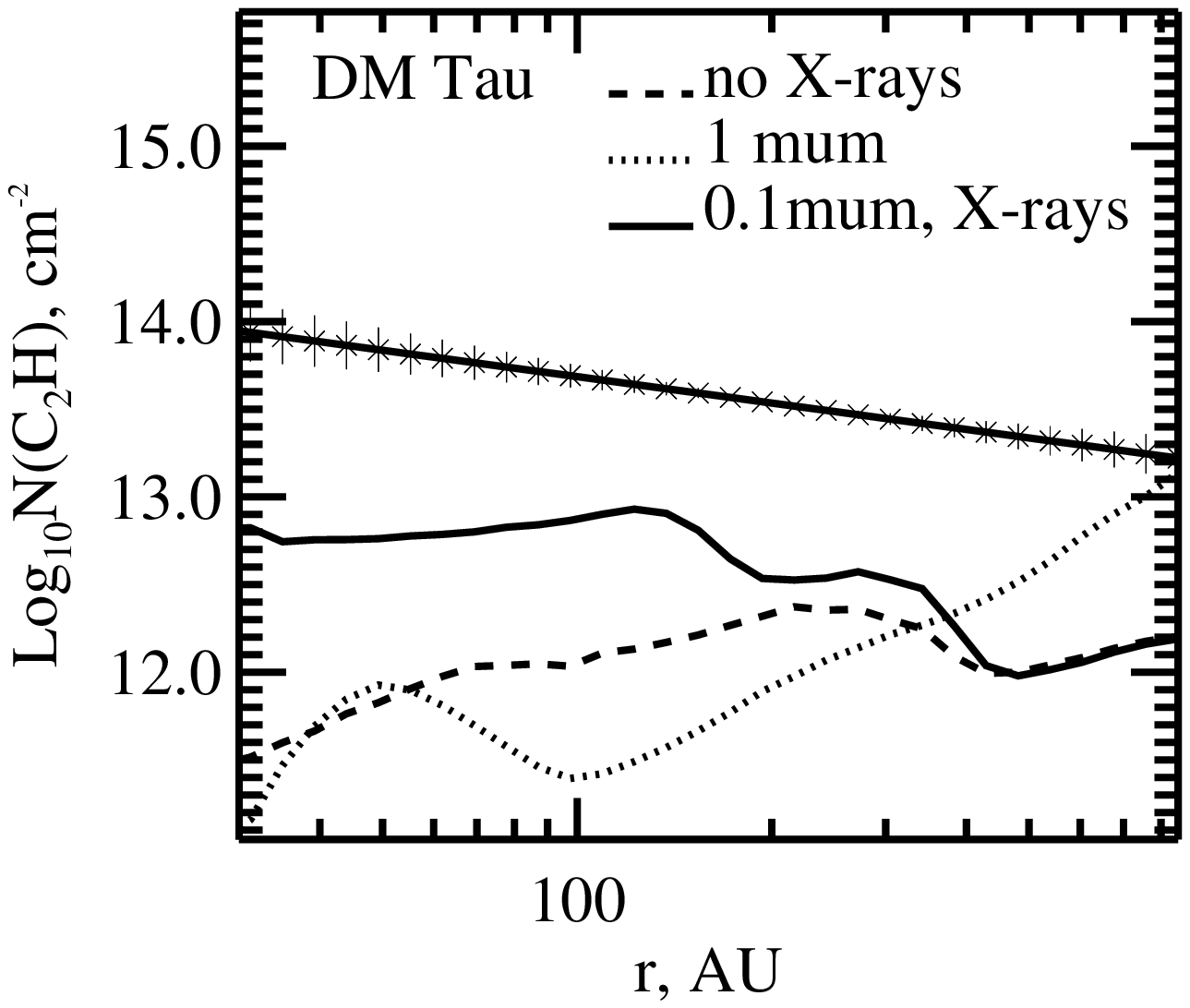}
\includegraphics[angle=90,width=0.31\textwidth]{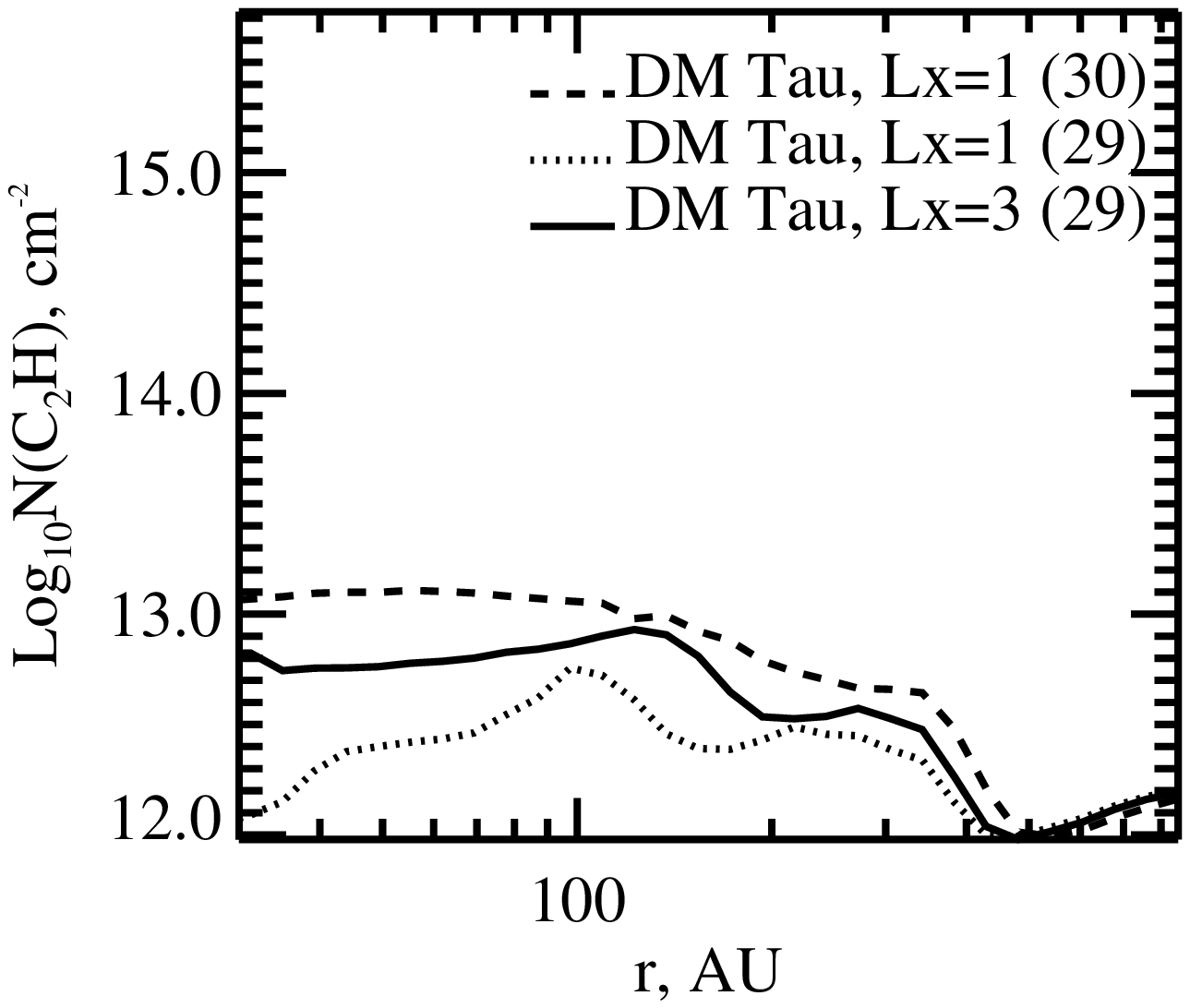}
\caption{ (Left) Radial distribution of the CCH vertical column densities in
the disks of DM Tau (solid line), LkCa 15 (dotted line), and MWC 480
(dashed line) at 5~Myr. The observed column densities and their
uncertainties are indicated by the lines with error bars. (Middle) Radial
distribution of the CCH vertical column densities in the DM Tau disk computed
with: (1) the standard model (solid line), (2) the model with bigger $1\mu$m
grains (dotted line), and (3) the model without X-rays (dashed line). (Right)
Radial distribution of the CCH vertical column densities in the DM Tau disk
computed with: (1) the $L_{\rm X}=3\,10^{29}$~erg\,s$^{-1}$ model (solid line),
(2) the $L_{\rm X}=10^{29}$~erg\,s$^{-1}$ model (dotted line), and (3) the
$L_{\rm X}=10^{30}$~erg\,s$^{-1}$ model (dashed line).}
\label{fig:cch_cd}
\end{figure}

The observed and modeled CCH column densities are compared in
Fig.~\ref{fig:cch_cd} (left panel). The CCH column densities in the DM~Tau disk
computed with the model without X-ray radiation and the model with bigger
$1\mu$m grains are depicted in Fig.~\ref{fig:cch_cd} (middle panel).
The influence of the adopted stellar X-ray luminosity on the CCH column
densities in the DM Tau disk is shown in the right panel.
As we mentioned above, the X-ray photons enhance CCH abundances, leading to
higher CCH column densities (by a factor of 3-10). In the model with
micron-sized grains CCH has high column density only in the very outer disk
region, at radii larger than $\sim 300$~AU, while it drops quickly in the inner
disk zone. Given the uncertainty of the modeled column densities \citep[a factor
of $\approx 3$, see][]{Wakelam_ea05,Vasyunin_ea08} and
observational uncertainties (see error-bars in Fig.~\ref{fig:cch_cd}),
our relatively simple steady-state disk
model is in a moderate agreement with the observed quantities.
In particular, the flat radial slope of the CCH column densities is
reproduced by the simulations. Note that the column density of CCH in MWC~480 is
lower compared to that of the DM~Tau and LkCa~15 disks, and a similar trend
exists in the observational data.

The effect of the stellar X-ray radiation field on the computed CCH column
densities is such that the lower the adopted X-ray luminosity, the lower the
resulting $N({\rm CCH})$, see Fig.~\ref{fig:cch_cd}, right panel. Apparently, a
transition from $L_{\rm X} = 10^{30}$~erg\,s$^{-1}$ to
$3\,10^{29}$~erg\,s$^{-1}$ is less prominent in the CCH column densities than
a transition from $L_{\rm X} = 3\,10^{29}$~erg\,s$^{-1}$ to
$10^{29}$~erg\,s$^{-1}$. Thus, indeed the ethynyl radical can serve as a
sensitive
probe of the impinging X-ray radiation field.
In general, it is the stellar UV luminosity that affects the evolution of CCH
globally, with X-ray radiation being
only important in inner disk regions ($r\la 50-200$~AU).

\section{Discussion}
\label{diss}
As we have shown in \citet{Schreyer_ea08}, the stellar UV luminosity can be a
crucial factor that makes the AB Aur system chemically poorer with respect to
the disk of DM Tau. The same trend is found here for C$_2$H and the
disks around MWC~480 and DM~Tau. However, we also show that X-ray
radiation plays an important role for high CCH concentration in the
inner regions of T Tau disks. Furthermore, the results of the steady-state disk
chemical model is in qualitative agreement with the CCH column densities
obtained by
$\chi^2$-fitting of the spectral maps. In contrast
to our CID1 paper \citep{Dutrey_ea07}, where such a model could not fit well
the radial slope of the N$_2$H$^+$ column densities, in the case of CCH the
flat distribution is obtained both from the analysis of the interferometric
data and chemical modeling.

\begin{figure}
\includegraphics[angle=90,width=0.95\textwidth]{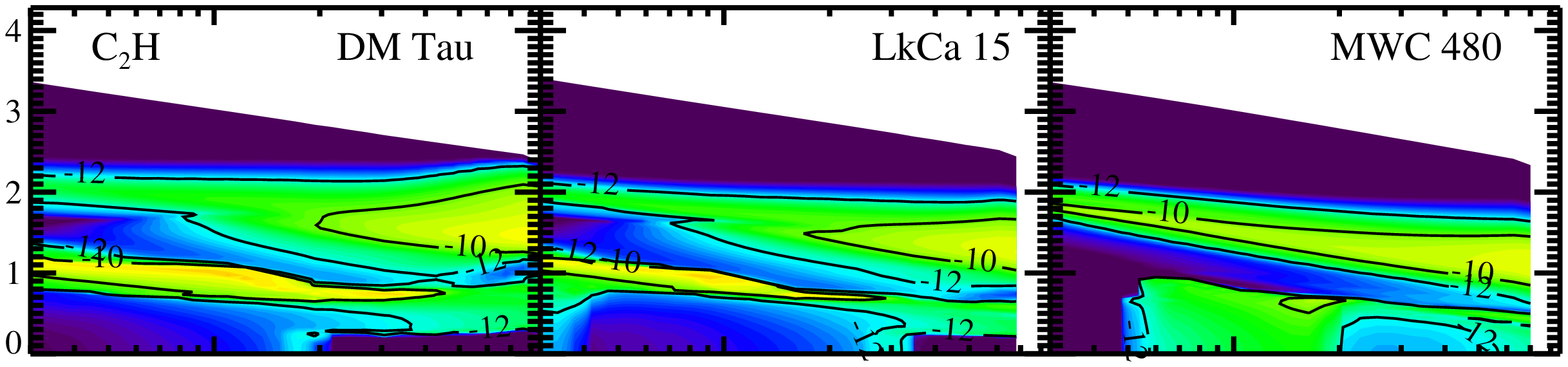}
\includegraphics[angle=90,width=0.95\textwidth]{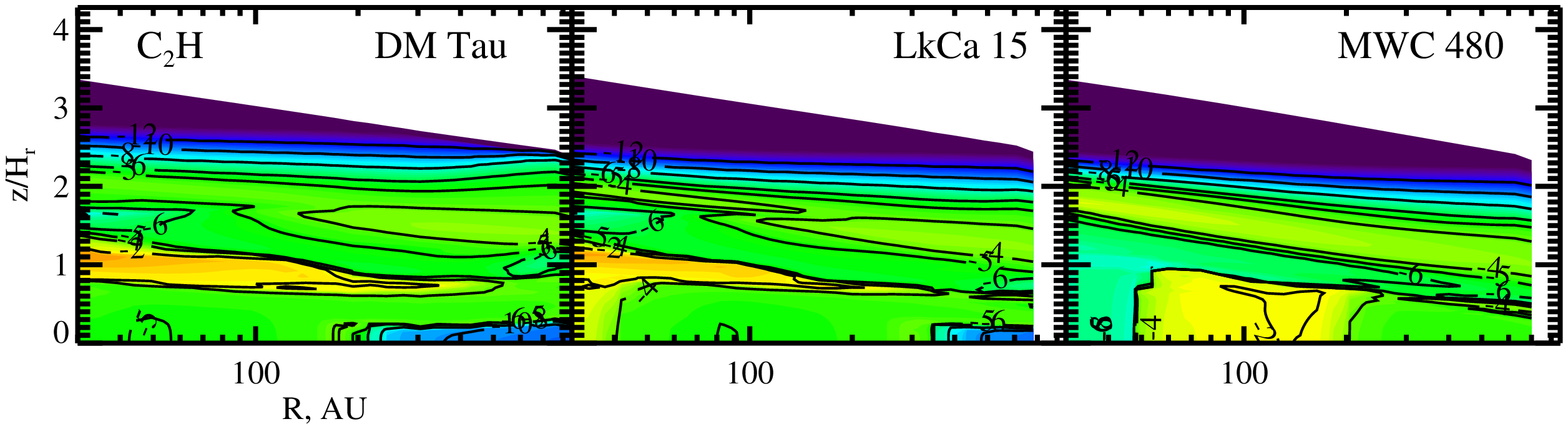}
\caption{(Top row; left to right) The CCH abundances (relative to the total
amount of hydrogen nuclei) in the DM Tau, LkCa 15, and MWC 480 disks at
5~Myr. The vertical axis is expressed in units of a pressure scale,
$H(r)=\sqrt(2)c_{\rm s}/\Omega$, where $c_{\rm s}$ is the sound speed in the
disk
atmosphere and $\Omega$ is Keplerian angular velocity. (Bottom row) The same as
in the top row but for the absolute CCH concentrations (cm$^{-3}$).}
\label{fig:cch_abunds}
\end{figure}

\begin{figure}
\includegraphics[angle=90,width=0.95\textwidth]{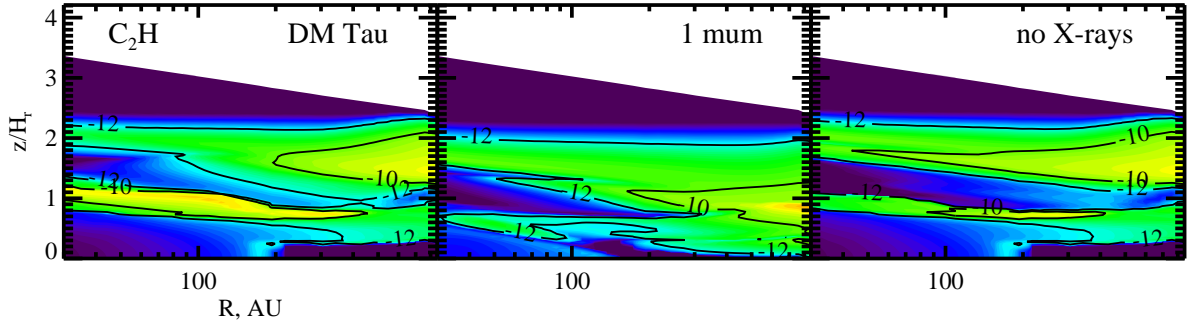}
\caption{(Left to right) The relative CCH abundances in the DM Tau disk
computed for 3 distinct models: (1) the standard model (top left panel,
Fig.~\ref{fig:cch_abunds}), (2) the model with bigger $1\mu$m grains,
and (3) the model without X-rays.}
\label{fig:cch_abunds_fact}
\end{figure}

Our chemical modeling predicts that two CCH molecular layers are
located at elevated disk heights, or in warm disk regions ($T\sim 35-75$~K).
This is inconsistent with the very low excitation temperature of the CCH
emission derived for DM Tau. The critical densities
for the CCH (1-0) and CCH(2-1) lines to be thermalized are
$2\,10^6$~cm$^{-3}$ and $7 \,10^6$~cm$^{-3}$, respectively. The two layers of
CCH have higher densities of $\sim 10^7-10^8$~cm$^{-3}$ in
all studied sources (compare Fig.~\ref{fig:disk_struc} and
Fig.~\ref{fig:cch_abunds}). Note that CCH layers are at higher densities
in the MWC~480 disk compared to those around LkCa~15 and DM Tau.
Hence, according to our model CCH(1-0) and (2-1) should be thermalized in all
considered sources, and sub-thermal excitation of these lines is unlikely.

\begin{figure}
\includegraphics[height=7cm]{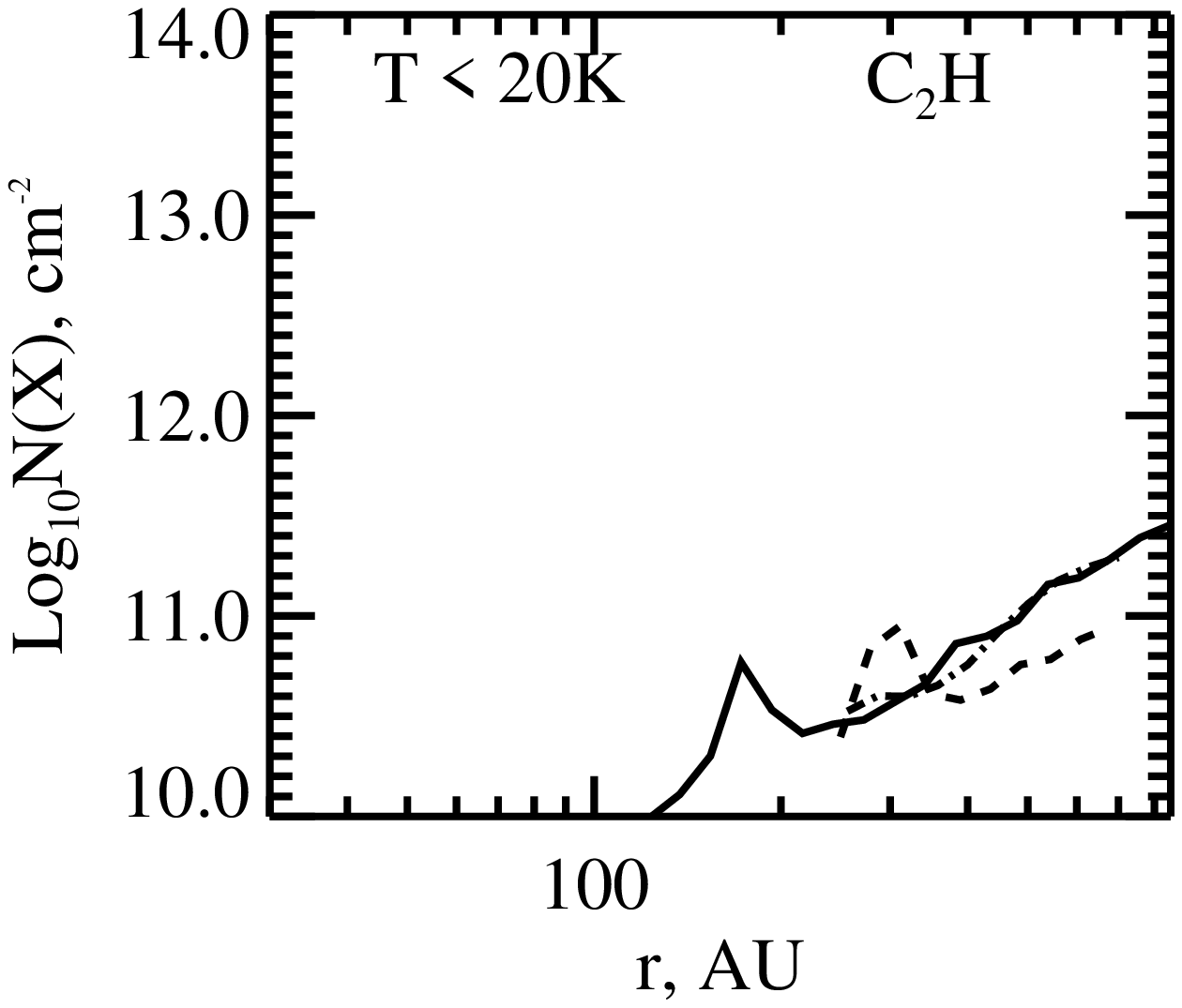}
\includegraphics[height=7cm]{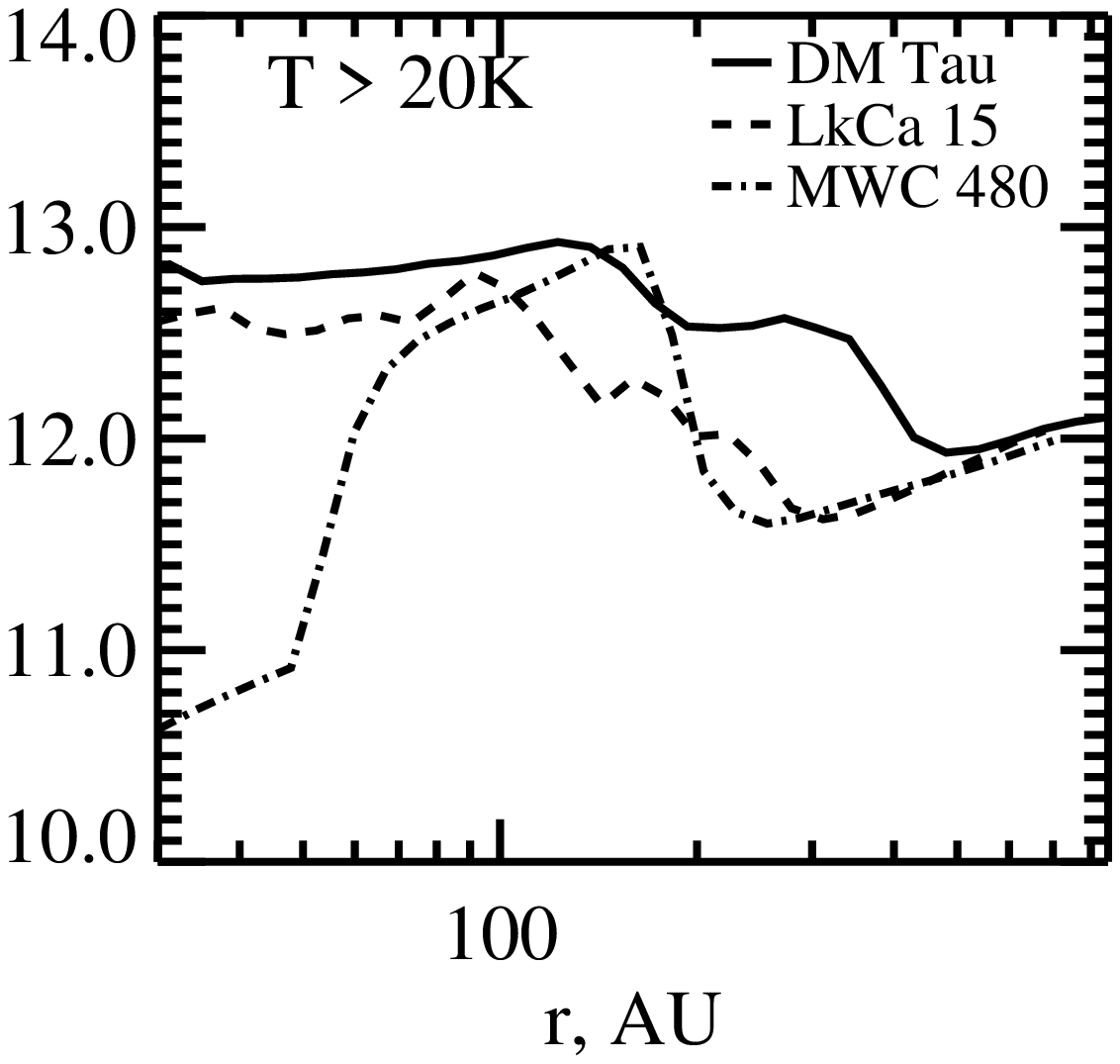}
\caption{Radial distribution of the CCH vertical column densities at 5~Myr in
the disk regions with $T<20$~K (left panel) and $T>20$~K (right panel) for DM
Tau (solid line), LkCa 15 (dotted line), and MWC 480 (dashed line). }
\label{fig:cold_cch}
\end{figure}

Which factors that are not taken into account in our simplified chemical
modeling could lead to a substantial concentration of CCH in the cold disk
midplane of DM Tau and LkCa~15? As we have found in \citet{Semenov_ea06}
\citep[and later confirmed by][]{Aikawa_07}, turbulent transport can enhance
the CO concentration in the DM Tau disk midplane such that $^{13}$CO (1-0) and
(2-1) lines are indeed ``cold'' ($T_{\rm ex} \la 15$~K). Later,
\citet{Hersant_ea08} have shown that the scattering of the stellar UV radiation
by PAHs and dust grains in upper disk atmosphere can redirect a portion of
these UV photons into the disk, allowing CO to efficiently desorb from dust
grains in the cold disk zone. Despite the fact that this assumption is
not considered in our model, where desorption of surface molecules
in the disk midplane is only due to the CRP-driven UV radiation and CRP
heating, a non-negligible amount of CCH exists in the outer disk midplane at
$T<20$~K, see Fig.~\ref{fig:cold_cch} (left panel). The CCH column density in
DM Tau at $T<20$~K has a maximum value of $6\,10^{11}$~cm$^{-2}$ at
$r=800$~AU, which rapidly decreases with radius. In warmer regions of the DM Tau
disk the CCH column density is about 5 times higher, $\approx
3\,10^{12}$~cm$^{-2}$, and has an almost flat profile. Thus, we expect that only
$\la 20\%$ of the CCH (1-0) and (2-1) line fluxes arise from the cold disk zone.

The amount of cold gases in disk midplanes can also be increased if gas-grain
interactions are ineffective such that freeze-out is delayed toward late
evolutionary times \citep[e.g.,][]{Aikawa_07}. We did several
test calculations with bigger grains of $1\mu$m and found that indeed neutral
species become more abundant in cold disk region. The results for molecular
chemistry in disks using a realistic grain evolutionary model will be reported
in a separate publication. In essence, in evolving disks substantial amount of
gas-phase molecules in very cold region can be produced and sustained by
moderate grain growth and sedimentation of micron-sized and bigger grains
toward the midplane, while a fraction of smaller grains and PAHs remains in the
disk atmosphere and provides enough opacities to keep the disk interior cool
\citep[see e.g.,][]{Aikawa_ea06,Jonkheid_ea07,Brauer_ea08,Birnstiel_ea09}.

Finally, in our disk model dust thermal balance is used to compute the
temperature distribution, which is assumed to be the same for the gas. In the
disk atmosphere gas will be hotter due to e.g. photoelectric heating of small
dust grains/PAHs by stellar/IS UV \& X-rays photons. We do not take this effect
into account. For CCH lines it is of no concern as derived low excitation
temperatures for the 1-0 and 2-1 transitions are indicative of a substantial
amount of ethynyl being located in the cold disk zone around the midplane. In
addition, our modeling results show that the majority of CCH is located at
scale heights of $\la 1$ for DM Tau and LkCa 15 and is even closer to the disk
midplane for MWC 480 (Fig.~\ref{fig:cch_abunds}, bottom row). When compared to
the corresponding disk density structures (Fig.~\ref{fig:disk_struc}), it
becomes apparent that the CCH molecular layers are within the disk regions
having particle densities of several $10^7$-$10^9$~cm$^{-3}$. These regions are
dense enough to allow effective coupling of gas and dust kinetic temperatures.

\section{Summary \& Conclusions}
\label{summary}
We observed protoplanetary disks around the Herbig Ae star MWC~480 and the T
Tauri stars LkCa~15 and DM~Tau in the (1-0) and (2-1) lines of CCH.
We detected and mapped these disks in the CCH lines, using the IRAM Plateau de
Bure Interferometer in the C- and D-configurations. Using the iterative
minimization technique, the column densities of ethynyl are derived, which
are lower in MWC~480 ($N({\rm CCH})\la 10^{13}$~cm$^{-2}$) than in DM Tau
and LkCa~15 ($N({\rm CCH})\sim 3\,10^{13}$~cm$^{-2}$). The derived excitation
temperature of CCH in DM Tau is very low, $T_{\rm ex}\sim 6$~K, which is along
with previous detections of cold gas in this object (CO, HCO$^+$, CH, and
HCN). The observed CCH column densities are compared
with the results of advanced chemical modeling, which is based on a
steady-state flared disk structure with a vertical temperature
gradient, and a gas-grain chemical network with surface reactions.
The CCH abundances in all disk models show two layers of high concentration.
Overall, the disk around the Herbig Ae star MWC~480 has less CCH, which we
explain
by the strong UV radiation field of this object, and the lack of X-ray
irradiation from the star.
The modeled CCH column densities are in qualitative agreement with the
observed values at an evolutionary time of a few million years. Altogether, our
measurements and modeling support the observational trend that disks around
Herbig Ae stars tend to be more
molecule deficient compared to those surrounding T Tauri stars due to stronger
UV irradiation and lower X-ray luminosities (affecting inner disk chemistry).
Our model reproduces the radial slope of the CCH column densities.
Yet it fails to explain the
low excitation temperature of the observed CCH lines as the modeled CCH
molecular layers are located in disk regions dense enough to thermalize both
transitions. The chemical simulations show that only
$\la 20\%$ of the CCH column densities are in the region of the DM Tau disk that
is colder than
20~K. Other mechanisms (stronger photodesorption, dynamic processes, grain
growth) are required to enhance the abundance of CCH in the cold disk
midplanes.
%
%
%
%
%

\acknowledgments
We acknowledge all the Plateau de Bure IRAM staff for
their help during the observations. DS acknowledges support by the
DeutscheForschungsGemeinschaft (DFG) through SPP 1385: ``The first ten million
years of the solar system - a planetary materials approach'' (SE 1962/1-1).
SG, AD, VW, FH, and VP are
financially supported by the French Program ``Physique Chimie du
Milieu Interstellaire'' (PCMI).  This research has made use of NASA's
Astrophysics Data System.


\end{document}